\documentclass[prd,eqsecnum,twocolumn,nofootinbib]{revtex4-2}

\usepackage{amsfonts}
\usepackage{mathtools}
\usepackage{amssymb}
\usepackage{amsthm}
\usepackage{bm}
\usepackage{dcolumn}
\usepackage{epsfig}
\usepackage{graphicx}
\usepackage{graphics}
\usepackage[latin1]{inputenc}
\usepackage{latexsym}
\usepackage{rotating}
\usepackage{xcolor}
\usepackage{gensymb}
\usepackage{makecell}

\newcommand{\f}[2]{\ensuremath{\frac{#1}{#2}}}

\begin{document}
\title{Gravitational waves from the late inspiral, transition, and plunge\\ of small-mass-ratio eccentric binaries}
\author{Devin R.\ Becker}
\affiliation{Department of Physics and MIT Kavli Institute, MIT, Cambridge, MA 02139 USA}
\author{Scott A.\ Hughes}
\affiliation{Department of Physics and MIT Kavli Institute, MIT, Cambridge, MA 02139 USA}
\author{Gaurav Khanna}
\affiliation{Department of Physics and Institute for AI \& Computational Research, University of Rhode Island, Kingston, RI 02881 USA}
\affiliation{Department of Physics, University of Massachusetts, Dartmouth, MA 02747 USA}
\affiliation{Center for Scientific Computing and Data Science Research, University of Massachusetts, Dartmouth, MA 02747 USA}

\begin{abstract}
Black hole binaries with small mass ratios will be important sources for the forthcoming Laser Interferometer Space Antenna (LISA) mission.  Models of such binaries also serve as useful tools for understanding the dynamics of compact binary systems and the gravitational waves they emit.  Using an eccentric Ori-Thorne procedure developed in previous work, we build worldlines that describe the full inspiral and plunge of a small body on an initially eccentric orbit of a Kerr black hole.  We now calculate the gravitational waves associated with these trajectories using a code that solves the Teukolsky equation in the time domain. The final cycles of these waveforms, the ringdown, contain a superposition of Kerr quasinormal modes followed by a power-law tail.  In this paper, we study how a binary's eccentricity and orbital anomaly angle affect the excitation of both quasinormal modes and late-time tails.  We find that the relative excitation of quasinormal modes varies in an important and interesting way with these parameters.  For some anomaly angles, the relative excitations of quasinormal modes are essentially indistinguishable from those excited in quasi-circular coalescences.  Consistent with other recent studies, we find that eccentricity tends to amplify the late-time power-law tail, though the amount of this amplification varies significantly with orbital anomaly.  We thus find that eccentricity has an important impact on the late-time coalescence waveform, but the interplay of eccentricity and orbit anomaly complicates this impact.
\end{abstract}
\maketitle

\section{Introduction}\label{sec:intro}
\subsection{Background, motivation, and past work}

Binary black hole mergers detected by the LIGO-Virgo-KAGRA (LVK) Collaboration are, to date, well described using circular coalescence waveform templates \cite{Abbott2023, LVK4}.  As the LVK gravitational wave (GW) catalog grows, an increasing number of events show evidence of more complicated orbital configurations \cite{RomeroShaw2022, RomeroShaw2020, Gupte2024, Wang2025, Abac2025}.  Looking beyond LVK, small-mass-ratio binaries detectable by future low-frequency GW experiments such as the Laser Interferometer Space Antenna (LISA) \cite{LISA2017, Babak2017, Colpi2024} are expected to enter the detection band in a variety of orbital geometries, with many formation mechanisms predicting substantial eccentricity, even during the merger \cite{Mancieri2025, Naoz2022, Bode2013, Bortolas2017, Bortolas2019, Mancieri2025cliff, Naoz2023, Rom2024}. 

A merging binary's coalescence geometry provides critical information about the formation channel which produced the system, a major point of interest for comparable mass systems \cite{Mapelli2021, Mandel2022, Abbott2016, Smith2025} and small-mass-ratio systems \cite{Sun2025} alike. Binaries with significant eccentricity are thought to originate from many-body interactions, such as dynamical encounters in dense stellar environments \cite{ Kremer2025, Zevin2021, Rodriguez2016, Zevin2019, Gondan2021, Miller2001} and three-body interactions \cite{DallAmico2024, Silsbee2017, ArcaSedda2021, Samsing2014, Antonini2017, Cocco2025A, Cocco2025B, Camilloni2023}. In contrast, aligned-spin systems with small or negligible eccentricity likely form in isolated stellar environments \cite{Tutukov1993, Belczynski2002, Belczynski2016}.  A diagnostic of eccentricity in the late waveform may serve to enhance the power of GW observations to constrain the relative importance of these different channels. 

In our previous work, Ref. \cite{becker2025}, hereafter BH25, we investigated the impact of orbital eccentricity on the late-time dynamics of black hole binaries in the small-mass-ratio regime.  Our model provides a framework for calculating a complete worldline for a small body spiraling into a Kerr black hole as the system emits GWs.  A complete worldline requires a prescription for the ``transition to plunge" that connects the slowly-evolving, adiabatic portion of the inspiral to the final plunge that rapidly takes the secondary beyond the primary's event horizon. Our work in BH25 calculates the transition by generalizing the Ori-Thorne procedure \cite{OriThorne2000}, originally developed for circular systems, to include eccentricity.  Refinement of the eccentric transition is still an active area of research\footnote{For additional work on the eccentric transition, see Refs.\ \cite{O_Shaughnessy2003, Sundararajan2008b, Sperhake_2018, Albanesi2023, Lhost2025, Honet2025, Faggioli2025} and for transition-to-plunge studies that are not restricted to eccentric systems, see Refs.\ \cite{OriThorne2000, Buonanno2000, Kesden2011, ApteHughes2019,  Burke2020, CompereKuchler2021, kuchler2024, Kuchler2025}.}.  Limitations of the BH25 model include a dependence on \textit{ad hoc} parameter choices, neglect of higher-order contributions from the gravitational self force \cite{Pound2022}, and relatively slow waveform generation.

Though these limitations must be addressed in order to model transition-to-plunge waveforms faithfully enough to be used to analyze detector data, we are confident that the worldlines generated using BH25 can be used to learn about the late-time dynamics of eccentric binaries and their associated GW signals.  Using trajectories from the framework of BH25, we build the source term for the Teukolsky equation \cite{Teukolsky1973}, and solve it with the time-domain code of Refs.\ \cite{Pranesh07, Pranesh08, Pranesh10, Zenginoglu2011, Field2021, Burko2016} to construct waveforms. Although this work formally applies to extreme mass ratio inspirals (EMRIs), black hole perturbation theory (BHPT) has proven to be a useful tool for understanding the relativistic two-body problem more generally \cite{Akcay2015, Wardell2023, LeTiec2011, VanDeMeent2020}.  We are hopeful that these results from the small-mass-ratio limit can be used to inform our understanding of eccentric waveform models over a wide range of mass ratios. 

We are especially interested in understanding whether eccentricity leaves a unique imprint on a binary's GW signal. While the impact of eccentricity on inspiral has been studied extensively, work to understand eccentricity's role in shaping the late-time \textit{ringdown} is ongoing.  During ringdown, the binary's merged remnant emits GWs as it settles to equilibrium \cite{Vishveshwara1970, Press1971, Teukolsky1973, Chandrasekhar1975}. First-order perturbation theory predicts that these waves have three main components: a prompt response, followed by a spectrum of quasinormal modes (QNMs), concluding with power-law tails. The prompt response is associated with the initial pulse propagating directly to the observer \cite{Leaver86, Nollert1999, Andersson1997}.  We do not examine the structure of the prompt response in this work, focusing our analysis on QNMs and power-law tails.

The ringdown is dominated by QNMs, which are modeled as damped sinusoids.  In the Regge-Wheeler-Zerilli (RWZ) formalism \cite{Regge57, Zerilli70, Martel2005, Nagar2005} QNMs are associated with poles of the RWZ Green's function \cite{DeAmicis2025}.  Observationally, QNMs offer the promise of testing general relativity: a Kerr black hole of mass $M$ and spin $a$ has a known spectrum of QNMs, each with its own frequency and damping time which depends only on $M$ and $a$.  Measuring two QNMs thus completely determines a merged remnant's mass and spin under the hypothesis that the remnant is a Kerr black hole; measuring more than two QNMs overdetermines the fit and allows one to test the Kerr hypothesis \cite{Echeverria1989, Dreyer2004, Berti2006, Berti2015, Berti2016, Thrane2017, Baibhav2018, Baibhav2019, Isi2019, Bustillo2021, Kamaretsos2012A, Kamaretsos2012B, Hughes2005, Cano2025, Berti2025}. The potential power of QNM measurement also goes beyond their frequencies: the relative amplitude of different QNMs is determined by the late-time geometry of the binary coalescence \cite{Pacilio2024, Lim2019, Hughes2019, Ma2021, Zhu2025, Nobili2025}. Previous work in both the small-mass-ratio limit \cite{Lim2019, Hughes2019, Nagni2025}, as well as with systems at less extreme mass ratios \cite{Ma2021, Zhu2025}, has shown that the inclination of the binary affects the relative excitation of QNMs. For quasi-circular and nonprecessing mergers in the equatorial plane, the $(\ell, m) = (2,2)$ fundamental QNM is likely the most dominant excited mode. For significantly misaligned coalescences, the $(2,1)$ or $(2,0)$ modes may dominate the spectrum.

This analysis focuses on equatorial orbits, investigating the correspondence between eccentricity and the excited QNM amplitudes. Quasinormal mode excitation from eccentric systems has also been investigated with the effective-one-body (EOB) framework \cite{Damour2009, Albanesi2023}, numerical relativity (NR) \cite{Carullo2024qnms, Ficarra2025} and first-principles models \cite{DeAmicis2025}.  Our conclusions in Sec.\ \ref{sec:Conclusion} outline possible points of synergy between our work and these.

QNMs exponentially decay, and their contribution to the ringdown becomes negligible after some finite number of cycles.  Once they have decayed sufficiently, the ringdown is dominated by ``Price" tails, in which the GW amplitude decays in a power-law with time \cite{Price72A, Price72B, DeWitt1960, Leaver86}.  These late-time tails are a consequence of gravitational radiation back scattering from the black hole potential.  In the RWZ framework, tails arise from a branch cut of the Green's function along the zero-frequency axis \cite{Ching1995,Casals2012, DeAmicis2024A}.  A great deal of work has gone into using tails to learn about the asymptotic structure of dynamical spacetimes and the relaxation of compact objects in general relativity \cite{Cunningham1978, Leaver86, Gundlach1994A, Gundlach1994B, Okuzumi2008, Burko1997, Barack1999, Bernuzzi2008, Burko2014, Zenginolu2014, Cardoso2024,Burko2011, Burko2004, Burko2008,  Krivan1999, Poisson2002, Burko2003, Barack1999B, Racz2011, Harms2013, Zenginoglu2010,Hod2000B}.  Recently, attention has focused on understanding the impact of a binary's coalescence geometry on tail excitation. Studies of nonspinning binaries with the RWZ formalism \cite{DeAmicis2024A, Albanesi2023}, and phenomenological work with BHPT that includes spinning binaries \cite{Islam2024, Islam2025} have all found that power-law tails are amplified by a system's eccentricity: tails take over the ringdown earlier and with larger amplitudes in eccentric systems than in their circular counterparts.

Tails have also been found in NR simulations of comparable-mass binaries: Ref.\ \cite{DeAmicis2024B} extracts tails from simulations of head-on collisions, Ref.\ \cite{Ma2025} observes tails in both head-on collisions and highly eccentric mergers, and Ref.\ \cite{Islam2024} points out possible evidence of tails in publicly available Rochester Institute of Technology simulations \cite{Healy2022} and data from the Simulating eXtreme Spacetimes (SXS) collaboration \cite{Boyle2019}\footnote{It's worth noting that Ref.\ \cite{Islam2024} was published before the most recent update to the SXS catalog, \cite{Scheel2025}, which includes many more eccentric simulations than in previous releases.}.  As discussed in Ref.\ \cite{Islam2024}, one must be quite careful when looking for tails in NR data.  First, publicly available NR simulations typically do not extend into the tail regime of the ringdown; second, simulations have only recently begun to cover the eccentric portion of parameter space where we expect tails to be strongest.  (Indeed, defining and controlling eccentricity in NR is fundamentally a challenge; see discussion in Refs.\ \cite{Habib2024, Knapp2025, Nee2025}.)  Future improvements to NR studies will likely reduce the necessity for these cautions.

\subsection{This paper: Ringdown from eccentric binary coalescences}

An important result of BH25 is that the late-time kinematics of an eccentric binary are sensitive to initial conditions.  As the secondary progresses through an eccentric inspiral, it radially oscillates between closest approach to the primary (periapsis) and furthest separation (apoapsis), with these defining orbit points adiabatically evolving due to GW backreaction.  We track where the system is in its orbital cycle using an orbital anomaly angle which grows monotically as inspiral proceeds.

BH25 shows that systems which are identical in all initial characteristics except anomaly angle can execute quite different transition-to-plunge dynamics.  For example, a system whose secondary passes periapsis and begins returning to apoapsis as the system enters the transition can complete a final radial oscillation before plunging; a system that is identical in all characteristics except the initial anomaly may enter the transition in such a way that the secondary momentarily ``whirls'' at periapsis before plunging.  The final kinematics of the ``whirling'' system may be barely distinguishable from the behavior of a quasi-circular inspiral.  Eccentric systems thus do not display the late-time universality previously shown to describe circular mergers \cite{OriThorne2000, Buonanno2000, ApteHughes2019}.  This sensitivity to the eccentric phase has also been observed in EOB transition-to-plunge trajectories in the test-mass limit \cite{Faggioli2025}, as well as NR simulations of symmetric binaries \cite{Nee2025}.

Not surprisingly, we find that this lack of universality carries over to ringdown excitation as well.  In particular, we find that the ringdown does {\it not} clearly encode a system's eccentricity.  Although the ringdown of an eccentric coalescence can be distinct from the ringdown of a quasi-circular merger, this typically holds only over a range of anomaly as the system enters transition.  In most cases we examine, we can find anomaly angles for which ringdown following eccentric inspiral is essentially indistinguishable from quasi-circular ringdown.  In particular, we find that systems which follow eccentric inspirals, but whirl in a quasi-circular fashion before plunging, produce ringdown spectra dominated by the $(\ell, m) = (2,2)$ mode, just as we see in quasi-circular equatorial coalescences.  If the anomaly is tuned such that the system does not exhibit this quasi-circular whirling, the $(2, 1)$ mode becomes significant, and can be the dominant mode.  This behavior also depends on the system's spin and eccentricity.  We explore these dependencies in Sec.\ \ref{sec:chiITqnm}.

The diversity of QNM behavior we find from systems that differ only in anomaly angle raises the question of which aspect of the system controls this behavior.  We have found there is a particularly strong correlation between the relative mode excitation and the radial velocity during the final plunge.  In particular, we find that QNM excitation depends very cleanly on the radial velocity $dr/d\tau$ as the secondary crosses the prograde equatorial light ring \cite{Khanna2017, Price2016}.  Systems which plunge directly from apoapsis tend to cross the light ring with high radial velocity; those which ``whirl'' in a quasi-circular manner near periapsis before plunging cross the light ring more slowly (kinematically quite similar to what is seen in a plunge from quasi-circular inspiral).  These results, discussed in detail in Sec.\ \ref{sec:drdtauITqnm}, indicate that the excitation of ringdown depends mostly on the detailed nature of the secondary's final plunge trajectory.


Though our main focus is on the excitation of QNMs, we also examine the behavior of the late-time tails which conclude these waveforms.  Focusing on the $h_{22}$ spherical mode, we find results broadly consistent with other recent studies of tails from eccentric mergers \cite{Islam2024, Islam2025, DeAmicis2024A, DeAmicis2024B, Albanesi2023}.  In particular, we find that, at fixed anomaly angle, increasing eccentricity $e$ increases the amplitude of the tail, causing it to become the dominant contributor to the waveform earlier.  The power-law index describing the decay of the tail remains insensitive to $e$: tails decay at the same rate, independent of eccentricity.  However, our analysis uncovers new phenomenology: the tails vary quite a bit as a function of orbital anomaly.

Reference \cite{Islam2025}, completed at the same time as the present paper and hereafter called IFK25, presents a far more comprehensive study of the behavior of tails in eccentric mergers, using the same code and methods as we use in this analysis.  IFK25 does not examine the influence of the anomaly angle, and uses EOB trajectories to construct their source terms.  However, they examine tails for modes with $(\ell, m) = (2,2)$, $(2,1)$, $(3,3)$, $(3,2)$, $(4,4)$, and $(4,3)$.  They also use sophisticated Bayesian estimation and MCMC methods to investigate correlations between model parameters, and the impact of the fitting interval on parameter inference.  This paper and IFK25 together highlight which system parameters influence the late-time tails of binary waveforms, and clarify how one can reliably extract this contribution to the waves from a binary model.

\subsection{Organization of this paper}

The remainder of this paper presents the details of our analysis.  We begin in Sec.\ \ref{sec:WLWFgeneration} by outlining our method for generating eccentric waveforms for small-mass-ratio systems.  We start by constructing the worldline that describes the secondary's inspiral, transition, and plunge into a Kerr black hole.  We describe this calculation at length in BH25; Sec.\ \ref{sec:WL} summarizes the relevant parts of this analysis, and defines key parameters used to describe the worldlines and their associated waveforms.  Section \ref{sec:WF} then summarizes how we use such a worldline to build the source term for the Teukolsky equation, as well as the computational tools we apply to solve the Teukolsky equation in the time domain.

The main focus of this analysis is the waveform's ringdown.  Section \ref{sec:Ringdown} describes the two ringdown components that we examine, QNMs and power-law tails.  We begin with an overview of Kerr QNMs, a discussion of the angular bases used in modeling them, and a description of the algorithm (which follows \cite{Lim2019}) that we employ to extract QNMs from our waveforms.  We then introduce power-law tails and describe how we model them.

We present results in Sec.\ \ref{sec:Results}, starting with QNMs.  We first show in Sec.\ \ref{sec:QNMoverview} some examples of QNM amplitudes from eccentric prograde and retrograde inspirals into Kerr black holes.  This overview gives an introductory sense for how the relative QNM excitation varies with eccentricity, highlighting the spread in mode amplitudes introduced by the initial radial phase.  Section \ref{sec:chiITqnm} explores this dependence on the radial phase in more detail, investigating different spins and a range of eccentricities.  We show that there exists a region of parameter space where the modes are extremely sensitive to the anomaly angle, changing significantly when the angle changes slightly.  The dependence on anomaly angle hints at the manner in which plunge geometry controls mode excitation, motivating our study of mode excitation as a function of the secondary's radial velocity in Sec.\ \ref{sec:drdtauITqnm}.

In Sec.\ \ref{sec:tailresults}, we shift focus to the power-law tails at the end of ringdown.  We begin by presenting a typical example of a late-time tail, then examine how tails depend on spin and eccentricity in Sec.\ \ref{TailsResultsEcc}.  This analysis shows that, with all other properties fixed, increasing eccentricity increases the strength of the tails, causing them to dominate ringdown earlier.  We then hold eccentricity constant to study the impact of the radial anomaly in Sec.\ \ref{sec:chiITtails}.  As was the case with the QNMs, we find that varying the radial anomaly has a significant effect on the overall amplitude of late-time tails.

Section \ref{sec:Conclusion} presents our conclusions.  Chief among them is that the detailed nature of ringdown in an eccentric coalescence is governed by the final kinematics of the binary as inspiral ends and the system enters its final plunge.  These final kinematics in turn are controlled by a combination of parameters, predominantly the binary's eccentricity and the anomaly angle which determines the radial motion as the secondary transitions from inspiral to plunge.  We also discuss ideas for extending this work, and ways to compare with related studies that do not focus on the extreme mass ratio limit.

We include three appendices of technical details which extend our discussion from elsewhere in this paper.  Appendix\ \ref{app:higherordermodes} expands the QNM analysis of Sec.\ \ref{sec:QNMresults} to include modes with $\ell > 2$.  The results for these modes are quite similar to what we find for $\ell = 2$; going to higher $\ell$ has no effect on our conclusions.  In Appendix \ref{app:chifurcation}, we study in greater detail how our results depend on the radial anomaly angle, exploring different mass ratios, as well as connections to related work.  Finally, in Appendix \ref{app:alphabeta} we show that the \textit{ad hoc} parameter choices in the BH25 worldline model do not consequentially impact the excitation of QNMs.

Throughout this work, we use geometrized units with $G = 1$ and $c = 1$. The larger black hole is described by the Kerr metric \cite{Kerr1963} in Boyer-Lindquist coordinates $\{t, r, \theta, \phi\}$ \cite{Boyer:1966qh}, with mass $M$ and spin parameter $a$.  The smaller body has mass $\mu$, and the mass ratio is $\eta = \mu/M$.  We restrict our analysis to orbits in the equatorial plane, where $\theta = \pi/2$. 

\section{Gravitational Waves From Small-Mass-Ratio Inspiral and Plunge for Eccentric Orbital Configurations}
\label{sec:WLWFgeneration}

\subsection{The inspiral, transition, and plunge worldline: How it is computed and key parameters describing it}
\label{sec:WL}

Our method for calculating an eccentric inspiral, transition, and plunge worldline for a small body spiraling into a Kerr black hole is discussed at length in BH25.  Here we give a brief synopsis of this framework, highlighting features that are key to our ringdown study:

\begin{enumerate}

    \item The small body begins on an eccentric geodesic in the equatorial plane of a Kerr black hole. We initiate the inspiral far enough from the last stable orbit (LSO) to model the dynamics as an \textit{adiabatic inspiral}, where the small body flows through a sequence of geodesic orbits of the large body's spacetime as the system emits GWs. The rate of flow through this sequence is determined by the flux of GWs at each geodesic, which we extract from precomputed radiation reaction data grids (see Refs.\ \cite{Chua2021, Katz2021, Speri2023, Hughes2021, Chapmanbird2025}).

    \item As the secondary approaches the LSO, the GW fluxes grow stronger, and the potential-like function which governs its radial motion grows flatter.  The backreaction of GW emission becomes so strong compared to the ``restoring force'' from the potential's curvature that the evolution can no longer be described as adiabatic.  At this point, the system has begun its \textit{transition to plunge}.  We then calculate the trajectory with our eccentric generalization of the procedure introduced by Ori and Thorne \cite{OriThorne2000}.  This description of the smaller body's motion through the transition relies on using the radial geodesic equation (see \cite{MTW}) with the orbit integrals $E$ and $L_z$ evolving due to backreaction.
    
    \item After crossing the LSO, the small body's motion is well approximated by a plunging Kerr geodesic with constant $E$ and $L_z$, which eventually freezes on the primary's event horizon (as parameterized by Boyer-Lindquist coordinate time).
    
\end{enumerate}

\begin{figure}
\centering
\includegraphics[scale = 0.43]{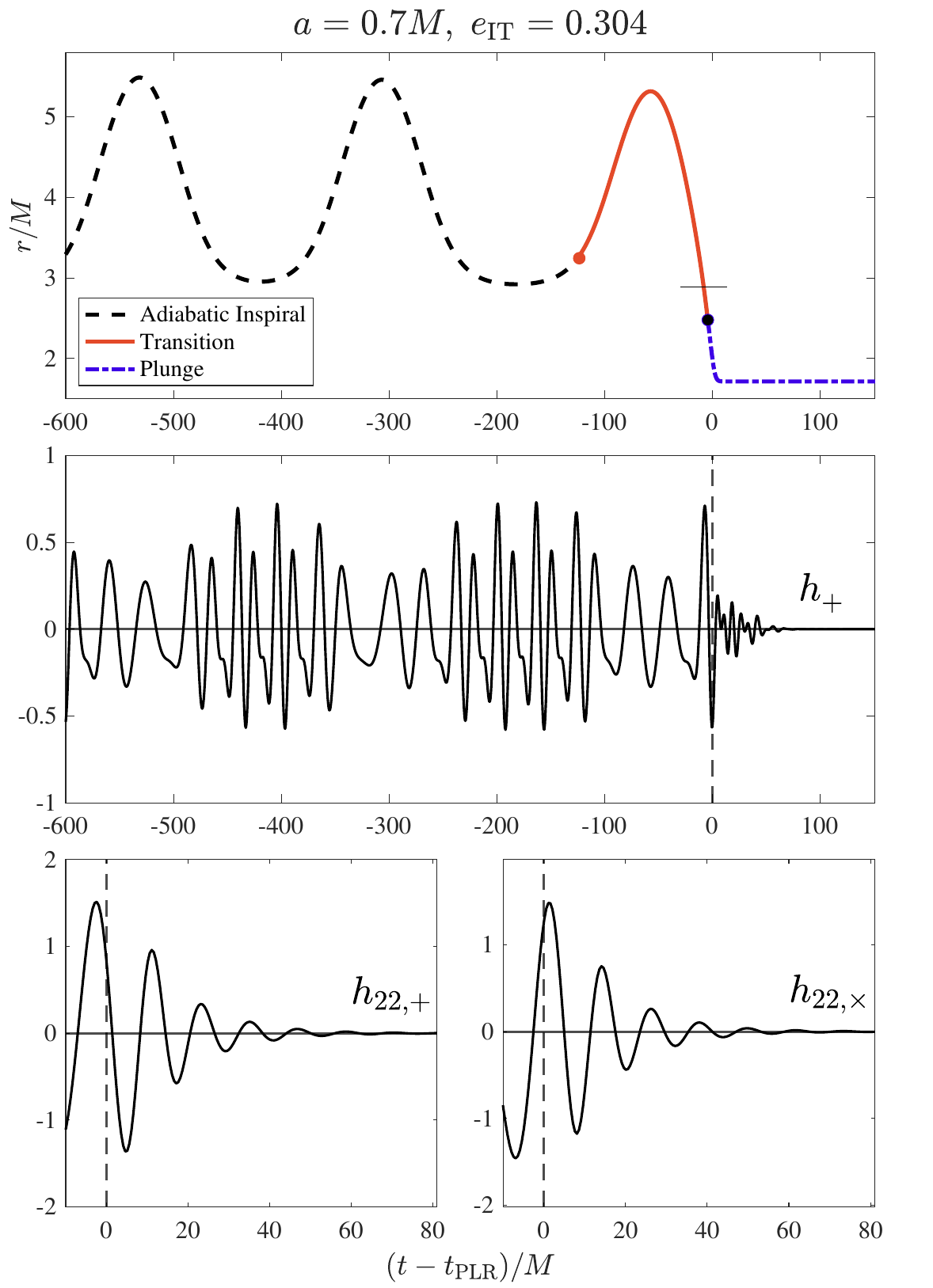}
\caption{Example of a prograde inspiral-transition-plunge worldline built using BH25 and its corresponding waveform. The binary has a mass ratio of $\eta =10^{-4}$ and spin $a = 0.7M$. Top panel shows orbital radius versus Boyer-Lindquist coordinate time.  The interval shown includes the final cycles of adiabatic inspiral, ending with an eccentricity of $e_{\rm IT} = 0.304$.  The secondary then completes a radial cycle in the transition, eventually crossing the LSO (the black bar on the plot). Past the LSO, we switch to a plunging geodesic, and the small body ultimately freezes to the horizon coordinate $r_{\rm H} =1.7414M$. Before freezing to the horizon, the secondary also crosses the prograde equatorial light ring at $t_{\rm PLR}$. Middle panel includes the plus-polarization waveform according to an ``edge-on" observer in the primary's equatorial plane. Bottom panels focus on the ringdown of the $\ell = m = 2$ component of the waveform: the bottom left panel is the plus polarization, and the bottom right panel is the cross polarization.}
\label{fig:WLWF}
\end{figure}
The top panel of Figure \ref{fig:WLWF} includes a prograde worldline calculated with this procedure. We plot the radial motion as a function of Boyer-Lindquist coordinate time $t$ for a small body's inspiral, transition, and plunge into a Kerr black hole with $a = 0.7M$, $\eta = 10^{-4}$ and $e_{\rm IT} = 0.304$.  (The subscript ``IT'' labels values at the end of adiabatic inspiral and the beginning of the transition.)  This system is confined to the equatorial plane of the primary, as is every system discussed in this paper.

Producing worldlines with this method requires two \textit{ad hoc} parameter choices.  The first, $\alpha_{\rm IT}$, determines when adiabatic inspiral ends and the transition begins.  It is defined as the ratio of the coordinate acceleration associated with geodesic motion at the inner turning point of the orbit with the acceleration associated with GW-driven backreaction; the exact definition is given in Sec.\ IIIB of BH25.
The second \textit{ad hoc} parameter, $\beta_{\rm TP}$, sets where transition ends and plunge begins.  This parameter sets a radius between the LSO and the event horizon at which we terminate the transition and treat the secondary's worldline as a plunging geodesic.  As $\beta_{\rm TP}\rightarrow 0$, the end of the transition approaches the LSO; as $\beta_{\rm TP}\rightarrow 1$, it approaches the horizon.

Although our transition and plunge worldlines depend on $\alpha_{\rm IT}$ and $\beta_{\rm TP}$, the ringdown does not strongly depend on these parameters.  We discuss this in detail in Appendix \ref{app:alphabeta}, examining how QNM excitation varies with $\alpha_{\rm IT}$ and $\beta_{\rm TP}$.  It is reassuring that these \textit{ad hoc} parameters do not significantly influence the ringdown sourced by a particular worldline, especially since BH25 noted that the worldline varies moderately with respect to $\alpha_{\rm IT}$ (though the dependence on $\beta_{\rm TP}$ is quite weak). Nonetheless, it remains dissatisfying that our prescription depends on these arbitrary values; a more fundamental understanding of how eccentric inspiral terminates would be extremely valuable.  Reference \cite{Lhost2025} discusses recent progress in describing the post-adiabatic eccentric inspiral, and is important progress in this vein.

Our transition and plunge trajectories also depend on initial conditions, namely the initial radial anomaly angle, $\chi_{r0}$. Along a geodesic, we can parametrize the radial motion by writing
\begin{equation}
    r = \frac{p}{1 + e\cos(\chi_r + \chi_{r0})}\;,
   \label{eq:rfun}
\end{equation}
where $e$ is the orbit's eccentricity and $p$ is its semi-latus rectum. When we initiate our adiabatic inspiral, we choose a starting value of the anomaly $\chi_{r0}$. As the secondary moves along the geodesic, the anomaly $\chi_r$ progresses from $0$ to $2\pi$. For more information on this particular parametrization, see \cite{Cutler1994}.

Trajectories that differ only by $\chi_{r0}$ will undergo identical, though out of phase, adiabatic inspirals.  However, when the secondary enters the transition, its motion is highly dependent on the radial phase, as this sets the transition's initial position and velocity.  For some values of $\chi_{r0}$, the small body will enter the transition and execute an additional radial oscillation before rapidly plunging into the primary's horizon; for others, it will ``whirl'' on a quasi-circular trajectory for some time before plunging.  An extensive discussion of the dependence of our transition and plunge trajectories on the relativistic anomaly is included in BH25.  In Secs.\ \ref{sec:chiITqnm} and \ref{sec:chiITtails}, we investigate the impact of the initial anomaly angle on the QNMs and tails excited by these systems.

Another useful quantity for assessing the impact of a system's late-time dynamics on its ringdown is the secondary's radial velocity as it crosses the prograde light ring (PLR). The location of the light ring depends only on the mass and spin of the primary, occupying the following radius in the equatorial plane:
\begin{equation}
    r_{\rm PLR} = 2M\left[1+\cos\left(\f{2}{3}\cos^{-1}\left[-\f{a}{M}\right]\right)\right]\;.
    \label{eq:pLR}
\end{equation}
Once the secondary reaches $r_{\rm PLR}$, we calculate its proper time radial velocity by evaluating the radial geodesic equation at $r_{\rm PLR}$ and with the values of $E$ and $L_z$ at this instant. In the equatorial limit, this velocity is
\begin{eqnarray}
    \left(\frac{dr}{d\tau}\right)_{r = r_{\rm PLR}} &=&  \frac{1}{r_{\rm PLR}^2}\left[[E(r_{\rm PLR}^2 + a^2) - aL_z]\right. 
    \nonumber \\ 
    &-& \left.\Delta_{\rm PLR}[r_{\rm PLR}^2 + (L_z - aE)^2]\right]^{1/2}\;,
    \label{eqn:drdtau} 
\end{eqnarray}
where $\Delta_{\rm PLR} = r_{\rm PLR}^2 - 2Mr_{\rm PLR} + a^2$.

In principle, the velocity at any moment during plunge could serve as a probe of the secondary's late-time dynamics.  We choose the PLR crossing in order to have a consistent comparison across systems, though we are also motivated by the role that particle dynamics at the light ring has held in ringdown studies (see, for example, Ref.\ \cite{Khanna2017}).  We present the relative QNM excitation as a function of the radial velocity at the PLR in Sec.\ \ref{sec:drdtauITqnm}.

\subsection{Solving the time-domain Teukolsky equation}\label{sec:WF}

Once we construct the inspiral-transition-plunge worldline by way of BH25, we use this worldline to build the source term for the time-domain Teukolsky equation \cite{Teukolsky1973}. The Teukolsky equation describes scalar, vector, and tensor field perturbations to Kerr spacetime, with tensor perturbations describing the radiative degrees of freedom of gravity.  We schematically write this equation
\begin{equation}
    \mathcal{D}^2\Psi = 4\pi \Sigma \mathcal{T}\;, 
\label{eq:teuk}\end{equation}
where $\mathcal{D}^2$ is a second-order differential operator (see \cite{Teukolsky1973} for its explicit form), $\Psi$ is related to the Newman-Penrose curvature scalar $\psi_4$ (see \cite{NewmanPenrose}) via $\Psi = r^4\psi_4$ for equatorial orbits, and $\mathcal{T}$ is the source term. We are particularly interested in $\psi_4$ because at future null infinity
\begin{equation}
    \psi_4 = \f{1}{2}\f{d^2}{dt^2}(h_+ - ih_\times)\;.\label{eqn:psi4dt2h}
\end{equation}
Hence, far from the source, $\psi_4$ encodes the emitted GWs. 

We compute the source $\mathcal{T}$ in Eq.\ (\ref{eq:teuk}) by using our worldline to build the secondary's energy-momentum tensor, which takes the following form in the equatorial limit:
\begin{equation}
    T_{\alpha \beta} = \mu\f{u_{\alpha}u_{\beta}}{r^2 u^t}\,\delta[r - r(t)]\,\delta[\phi - \phi(t)]\,\delta[\theta - \pi/2]\;,
\label{eqn:stressenergy}
\end{equation}
where $u^{\alpha}$ is the secondary's 4-velocity along its worldline.  We then project $T_{\alpha \beta}$ onto certain legs of the Kinnersley tetrad and operate on the resulting expression with a second-order differential operator to get $\mathcal{T}$; see \cite{Teukolsky1973}. 

Once we have $\mathcal{T}$, we solve Eq.\ (\ref{eq:teuk}) in the time domain using methods outlined in Refs.\ \cite{Pranesh07, Pranesh08, Pranesh10, Zenginoglu2011, Field2021, Burko2016}. The waveforms in the middle and bottom panels of Figure \ref{fig:WLWF} were calculated with this procedure, using the worldline in the top panel as the source. The middle panel gives the full waveform from our numerical code; the bottom panels focus on the ringdown, showing contributions from the plus and cross polarizations of the $(\ell, m) = (2,2)$ spherical mode.  Our code decomposes waveforms as 
\begin{equation}
    h(t) = \sum_{\ell,m} h_{\ell m}(t) _{-2}Y_{\ell m}(\theta, \phi)\;, 
\label{eqn:hlmN}
\end{equation}
where $h_{\ell m}(t) = h_{\ell m, +}(t) - ih_{\ell m, \times}(t)$, and $(\ell, m)$ is the spherical mode index. The functions $_{-2}Y_{\ell m}$ are $-2$ spin-weighted spherical harmonics \cite{Goldberg1967}.  (We discuss some issues related to the spherical mode decomposition in Sec.\ \ref{sec:YlmSkm}.)

For the QNM-dominated portion of the ringdown, the GW strain is well above the numerical noise floor of the time-domain Teukolsky solver discussed in Refs.\ \cite{Pranesh07, Pranesh08, Pranesh10, Zenginoglu2011, Field2021}.  Power-law tails, however, contribute at much lower amplitudes, especially for small eccentricity systems.  To perform more extensive studies of the tail behavior in both our work and IFK25, this time-domain solver has been adjusted to resolve waveform features at much lower amplitudes, by, for example, upgrading its infrastructure to support higher-precision floating-point arithmetic \cite{Burko2016}.

\section{Describing a Waveform's Ringdown Content}\label{sec:Ringdown}

The methods outlined in Sec.\ \ref{sec:WLWFgeneration} produce the time-domain waveform sourced by the inspiral, transition, and plunge of a small body into a Kerr black hole. We now shift our focus to the waveform's final cycles --- the ringdown.
First-order perturbation theory predicts that the ringdown includes contributions from both QNMs and power-law tails. In this section, we briefly define these ringdown components and summarize the fitting procedures we use to extract them from our waveforms.

\subsection{Quasinormal Modes}
\label{sec:QNMs}

In its early stages, the ringdown can be modeled as a linear superposition of QNMs. To understand QNMs more concretely, we first return to the Teukolsky equation (\ref{eq:teuk}).  Teukolsky showed \cite{Teukolsky1973} that Eq.\ (\ref{eq:teuk}) separates when decomposed in the frequency domain, resulting in ordinary differential equations. The solution for a given mode with frequency $\sigma\in\mathbb{C}$ is then given by 
\begin{equation}
    \Psi = \sum_{k=2}^{\infty}\sum_{m = -k}^k\sum_{n = 0}^\infty {}_{s}R_{k mn}^{a\sigma}(r) _{s}S_{k mn}^{a\sigma}(\theta, \phi)e^{-i\sigma t}\;.
\label{eq:psiQNM}
\end{equation}
By virtue of separability, the radial dependence of the solution is entirely described by $_{s}R_{k mn}^{a\sigma}(r)$, and the angular dependence by $_{s}S_{k mn}^{a\sigma}(\theta, \phi)$. The functions $_{s}S_{k mn}^{a\sigma}$ are spin-weighted \textit{spheroidal} harmonics. When $a=0$, these simplify to spin-weighted \textit{spherical} harmonics. We discuss the interplay between the spherical and spheroidal harmonic bases in the following subsection. For more information on spherical and spheroidal harmonics, as well as the radial function $_{s}R_{k mn}^{a\sigma}$, see Ref.\ \cite{Berti2009}. The indices $k$ and $m$ give the multipole and azimuthal numbers of the spheroidal harmonics, respectively, and the index $n$ is the overtone number \cite{Berti2009, Press1973}. Note that $\{r, \theta, \phi\}$ in this context is the radius and angular position at which the field is measured, not the binary's orbit position.

For certain frequencies, the modes in Eq.\ (\ref{eq:psiQNM}) satisfy physical boundary conditions: they describe radiation that is purely ingoing at the horizon and purely outgoing at null infinity. These solutions are the QNMs of the black hole, whose frequencies we denote $\sigma_{k m n}$. We can write $\sigma_{k m n}$ as 
\begin{equation}
    \sigma_{k m n} = \omega_{k m n} - i/\tau_{k m n}\;,  
\end{equation}
where $\omega_{k m n}$ is the oscillation frequency of the QNM, and $\tau_{k m n}$ is the mode's damping time.

So long as the product of the merger is a Kerr black hole, the excited QNM frequencies depend only on the mass $M$ and spin $a$ of the merger remnant. In the small-mass-ratio limit, we take both $M$ and $a$ to be the mass and spin of the primary. For a given $M$ and $a$, we pull $\omega_{k m n}$ and $\tau_{k m n}$ from the {\tt qnm} Python package \cite{Stein2019}, though tools for calculating both $\omega_{k m n}$ and $\tau_{k m n}$ and tables reporting their values are also included in \cite{Berti2009, Berti2006}.

As a consequence of the Teukolsky equation's symmetry, QNMs from Kerr black holes come in pairs: for a given set $(k, m; a, M)$, the eigenvalue problem admits two solutions: $_{s}S_{k mn}^{a\sigma}(\theta, \phi)$, with a positive oscillation frequency, and $_{s}S_{k-mn}^{a\sigma}(\pi - \theta, \phi)^*$, with a negative oscillation frequency and a different damping time (where $^*$ represents complex conjugation).  The latter contributions are called ``mirror modes'' (see \cite{Berti2006, Dhani2021}). Taking into account contributions from both regular modes and their mirrors, and specializing to spin weight $-2$, we follow $\cite{Berti2006, Lim2019, Hughes2019}$ and write the QNM contribution to the waveform as
\begin{widetext}
\begin{equation}
     h^{\rm QNM}(t) = \sum_{kmn}[\mathcal{A}_{kmn}e^{-i[\sigma_{kmn}(t-t_0)-\phi_{kmn}]}{}_{-2}S_{kmn}^{a\sigma_{kmn}}(\theta, \phi) + \mathcal{A}'_{kmn}e^{i[\sigma^*_{kmn}(t-t_0)+\phi'_{kmn}]}{}_{-2}S_{kmn}^{a\sigma_{kmn}}(\pi-\theta,  \phi)^*]\;.
    \label{eqn:mirrormodeexp}
\end{equation}
\end{widetext}
The constants $\{\mathcal{A}_{kmn}, \mathcal{A}'_{kmn}, \phi_{kmn}, \phi'_{kmn}\}$ are the amplitudes and phases associated with the regular mode (unprimed) and its mirror mode (primed) for each $(k,m,n)$. 

We assume the waveform is ringdown dominated when $t\geq t_0$.  Note that there is no unambiguous definition for $t_0$, especially in the small-mass-ratio regime \cite{Dorband2006, BertiCardoso2006}.  Previous work has shown that the extracted QNM amplitudes and phases depend on $t_0$; however, the {\it relative} modes are largely insensitive to this parameter \cite{Dorband2006}.  Studies focusing on comparable mass systems often use peaks in the waveform's orbital frequency \cite{Ficarra2025} or in the GW amplitude \cite{Buonanno2007, London2014, London2016,London2020} to mark the start of ringdown; more recent work has made progress systematically sampling different start times near these regions (cf.\ \cite{Mitman2025, Cheung2024, Zhu2024, Baibhav2023, Mitman2023, Cheung2023, Nee2023, Bhagwat2020}).  However, EMRI systems do not always have a clear peak in either frequency or amplitude.  We instead assume ringdown has begun once the secondary crosses the prograde equatorial photon ring \cite{Khanna2017, Price2016}.

The magnitude of each mode also depends on $\mu/D$, the ratio of the secondary's mass to the distance of the system: $\mathcal{A}_{kmn}\propto \mu/D$.  We set $\mu/D = 1$; results can be simply rescaled by introducing the factor $\mu/D$.


\subsubsection{Spherical and spheroidal expansions}
\label{sec:YlmSkm}

As illustrated by Eq.\ (\ref{eqn:hlmN}), our code that solves the Teukolsky equation decomposes waveforms in a spherical harmonic basis.  Since we fit numerical data projected onto a spherical basis to a QNM model in a spheroidal basis, we must take mode mixing into account \cite{Kelly2013, Berti2014, Cook2014, Taracchini2014, London2019}. Mode mixing occurs because basis functions with the same azimuthal index $m$ overlap: 
\begin{equation}
_{-2}S_{kmn}^{a\sigma_{\ell m n}}(\theta, \phi) = \sum_{\ell} \mu^*_{m\ell kn}(a\sigma_{\ell mn}) _{-2}Y_{\ell m}(\theta, \phi)\;, 
    \label{eqn:modemix}
\end{equation}
where $k$ is the spheroidal mode index and $\ell$ is a spherical mode index.  The overlap coefficients $\mu_{m\ell kn}$ are included in the Python \texttt{qnm} package; they can also be calculated using the algorithm discussed in Appendix A of Ref.\ \cite{Hughes2000} (see also Ref.\ \cite{Hughes2000Erratum}), and are tabulated in \cite{Berti2014}.

To fold mode mixing into our waveform analysis, we first equate the left-hand side of Eq.\ (\ref{eqn:hlmN}), our numerical data, to the right-hand side of Eq.\ (\ref{eqn:mirrormodeexp}), the QNM contribution to the ringdown. We then multiply both sides of the resultant equation by $ _{-2}Y_{\ell m}^*(\theta, \phi)$, integrate over the sphere, and apply Eq.\ (\ref{eqn:modemix}). We then find
\begin{equation}
    h_{\ell m}(t) = \sum_{k = k_{\rm min}}^\infty \sum_{n = 0}^\infty [a_{m\ell k n}(t)\mathcal{C}_{kmn} + a'_{-m\ell kn}(t)\mathcal{C}'_{k-mn}]\;, 
    \label{eqn:hQNMfitinf}
\end{equation}
where $k_{\rm min} = \textrm{max}(2, |m|)$. The constants $\mathcal{C}_{kmn}$ and $\mathcal{C}'_{k-mn}$ encode the QNM amplitudes and phases: 
\begin{equation}
    \mathcal{C}_{kmn} \equiv \mathcal{A}_{kmn}e^{i\phi_{km}}, \quad \mathcal{C}'_{kmn} \equiv \mathcal{A}'_{kmn}e^{i\phi'_{km}}\;,
\end{equation}
and the overlap coefficients are absorbed into the time-dependent coefficients:
\begin{eqnarray}
    a_{m\ell kn}(t) &=& \mu^*_{m\ell kn}(a\sigma_{kmn})e^{-i\sigma_{kmn}(t-t_0)}, \nonumber\\
    a'_{m\ell kn}(t) &=& (-1)^{\ell}\mu_{m\ell kn}(a\sigma_{kmn})e^{i\sigma^*_{kmn}(t-t_0)}\;.
\end{eqnarray}
For additional details of the derivation of Eq.\ (\ref{eqn:hQNMfitinf}), see Ref.\ \cite{Lim2019}.

\subsubsection{Mode extraction: A quick synopsis}
\label{sec:Limalg}

The code we use to extract QNMs from our waveforms was developed for the analysis carried out in $\cite{Lim2019, Hughes2019}$. Here we summarize the key features of this algorithm; for a more comprehensive discussion, see Sec.\ IIIC of $\cite{Lim2019}$. We begin by describing the basic features of the code, as well as the simplifications it makes to the general QNM picture introduced in Sec.\ \ref{sec:QNMs}. 

In contrast to similar studies carried out on comparable mass binaries, by working in the small-mass-ratio limit, we know the mass and spin of the remnant black hole \textit{a priori}. Since the mass and spin of the merged remnant determine the expected spectrum of QNMs, the algorithm assumes the ringdown contains those Kerr QNM frequencies and extracts their amplitudes and phases.  We also neglect overtone modes, i.e.\ QNMs for $n \geq 1$.  As $n$ increases, the damping time of the mode $\tau_{kmn}$ decreases \cite{Leaver86}.  We assume that by $t - t_0 \gtrsim 25M$, the ringdown is dominated by the fundamental QNMs.  The excitation of overtones is an important and interesting topic, but not one that we explore in this work.


In principle, the sum over $k$ in Eq.\ (\ref{eqn:hQNMfitinf}) is taken to infinity. In practice, the overlap coefficients peak when $k = \ell$ and then quickly decrease.  We thus truncate the sum at a finite value $k_{\rm max} = \ell + K_{\ell}$, where $K_{\ell}$ is determined by a procedure that we describe momentarily. 

These simplifications to our sums over $k$ and $n$ allow us to rewrite Eq.\ (\ref{eqn:hQNMfitinf}) as 
\begin{equation}
h_{\ell m}(t) = \sum_{k = k_{\rm min}}^{\ell + K_{\ell}}[a_{m\ell k 0}(t)\mathcal{C}_{km0} + a'_{-m\ell k0}(t)\mathcal{C}'_{k-m0}]\;.
    \label{eqn:hQNMfit}
\end{equation}
Using our our numerical data, $h_{\ell m}$, we calculate $\mathcal{C}_{km0}$ and $\mathcal{C}'_{km0}$ with the iterative fitting routine of \cite{Lim2019}.  The basic structure of this algorithm is as follows:

\begin{enumerate}

    \item Begin by picking some number $\mathcal{N}$ of spherical modes $(\ell_i, m)$ to incorporate into the fit.  Set $k_{\rm \min} = \ell_1 = \textrm{max}(2, |m|)$. The choice of $\mathcal{N}$ determines the number of mixed modes included beyond $\ell = |m|$, thus setting $K_{\ell}$. For example, if we are fitting $h_{22}$ spherical data, choosing $\mathcal{N} = 1$ would only include the $\ell = 2$ modes, while choosing $\mathcal{N} = 3$ includes the $\ell = 2$ modes, as well as the $\ell = 3$ and $\ell = 4$ modes.
    
    \item Evaluate Eq.\ (\ref{eqn:hQNMfit}) and its time derivative at some $t>t_0$. Solve the resulting $2\mathcal{N}$ linear equations for the mode amplitudes $\mathcal{C}_{km0}$ and $\mathcal{C}'_{k-m0}$.
    
    \item Repeat the previous step at multiple times in the QNM-dominated regime, again solving for $\mathcal{C}_{km0}$ and $\mathcal{C}'_{k-m0}$. This tests the consistency of the algorithm: if the values found for $\mathcal{C}_{km0}$ and $\mathcal{C}'_{k-m0}$ are stable, then the ringdown model is consistent with the data $h_{\ell m}$.  If the values of $\mathcal{C}_{km0}$ and $\mathcal{C}'_{k-m0}$ are not stable (e.g., their values oscillate over time), then the ringdown model is inadequate.  The model may just need more spheroidal modes, so increase $\mathcal{N}$ and repeat the procedure.  (It may also be the case that the radiation is not dominated by QNMs; no stable solution will emerge in this case.)
    
    \item Once $\mathcal{C}_{km0}$ and $\mathcal{C}_{k-m0}$ stabilize, calculate a moving average with fixed $\Delta t$.  Set $\overline{\mathcal{C}}_{km0}$ and $\overline{\mathcal{C}}'_{k-m0}$ to the average with the least variance. The ringdown model is then given by
    \begin{equation}
    h_{\ell m}^{\rm RD}(t) = \sum_{k = k_{\rm min}}^{\ell + K_{\ell}} [a_{m\ell k0}(t)\overline{\mathcal{C}}_{km0}+ a'_{-m\ell k0}(t)\overline{\mathcal{C}}'_{k-m0}]\;.
    \end{equation}
    \end{enumerate}
We point the reader to Ref.\ \cite{Lim2019} for more of the computational details of this fitting procedure, as well as consistency checks between this algorithm and previous work.

\subsection{Power-law tails}
\label{sec:tails}

At sufficiently late times, the ringdown is composed of slowly decaying, power-law tails. In terms of the Teukolsky function $\psi_4$, the tail contribution to the ringdown can be modeled as
\begin{equation}
    [\psi^{\rm tail}_4]_{\ell m}(t) = \mathcal{A}_{\psi\textrm{tail}}^{\ell m}(t+c_{\rm tail}^{\ell m})^{p_{\psi\rm tail}^{\ell m}}e^{i\phi_{\rm tail}^{\ell m}}\;, \label{eqn:psi4tails}
\end{equation}
where $\psi_4$ is projected onto the same basis of spherical harmonics we use to describe $h_{\ell m}$. The parameter $\mathcal{A}_{\psi \textrm{tail}}^{\ell m}$ sets
the amplitude of the tail when it begins dominating the waveform, $c_{\textrm {tail}}^{\ell m}$ is a time-shift parameter, and $p_{\psi \textrm {tail}}^{\ell m}$ is the power-law index.  We expect a power-law index of $p_{\psi\rm tail}^{\ell m} = -(\ell +4)$ \cite{Barack1999, Hod2000A} to describe the decay of $[\psi_4]_{\ell m}$.  This differs from the power-law index of tails in $h_{\ell m}$, for which we expect $p_{h\rm tail}^{\ell m} = -(\ell +2)$.  As we explain in more detail below, we extract tails from $[\psi_4]_{\ell m}$ rather than $h_{\ell m}$ to avoid numerical noise effects which can mask tails in our data.  We focus on the $\ell = m = 2$ harmonic mode, for which we expect $p_{\psi\rm tail}^{22} = - 6$. Tails from higher-order spherical modes decay more rapidly, making them difficult to isolate in numerical data. Such contributions, as well as tails from $\ell = 2$ modes with $m\neq2$, are studied at length in IFK25.

In both our study and that presented in IFK25, the waveforms are extracted at future null infinity.  The code directly computes $[\psi_4]_{\ell m}$.  By Eq.\ (\ref{eqn:psi4dt2h}), $h_{\ell m}$ can then be found by twice integrating $[\psi_4]_{\ell m}$.  This double integration tends to amplify numerical noise.  Although this noise is ``harmless'' for much of our analysis, it is extremely important when attempting to extract tails, especially for low eccentricities for which the amplitude tends to be exceedingly small.  (Past work \cite{Islam2024} overcame this limitation by studying systems with eccentricity $e > 0.8$.)  Because they are simply related by time derivatives (or integrals), either $[\psi_4]_{\ell m}$ or $h_{\ell m}$ can be used to diagnose the properties of tails. One must, however, bear in mind which particular field is used in a given analysis to make sure that one compares like with like between different analyses.

\subsubsection{Tail extraction: A quick synopsis}

We fit the tails using \texttt{gwtails}, a Python package developed in Ref.\ \cite{Islam2024} and expanded for IFK25. Our study applies the portion of \texttt{gwtails} that uses \texttt{scipy.curve-fit} for a quick estimate of the parameters $\{A_{\psi \textrm{tail}}, c_{\rm tail}, p_{\psi\textrm {tail}}\}$ with nonlinear least-squares optimization.  This is sufficient for our initial study.  A deeper analysis of late-time tails, such as that given in IFK25, requires a more complete characterization of these parameters, examining their correlations and dependence on the chosen fitting interval.  This is done with the update to \texttt{gwtails} made with IFK25, which introduces Bayesian inference and MCMC methods to study the fit parameters more thoroughly. 

As the QNM contribution decays, the waveform exhibits brief oscillations before the tail dominates. These oscillations arise from QNMs mixing with tails and should be omitted both from the QNM extraction regime and the region considered for tail fits \cite{Albanesi2023, DeAmicis2024A, Islam2024, DeAmicis2024B, Ma2025}.  We consequently begin fitting the power-law tail after the oscillations vanish.  This usually occurs when $t - t_{\rm PLR}\gtrsim 300M - 500M$, though the precise start time varies with eccentricity and the late-time anomaly angle.  Ideally, we would start fitting tails even later, but the ringdown data we analyze only extend to $t - t_{\rm PLR}\approx 1000M$, and good fit convergence typically requires a time range larger than about $300M$.  The IFK25 analysis calculates their waveforms to $9000M$ after the merger, so they are able to provide a more complete analysis of how the fitting interval affects tail inference.

\section{Results}
\label{sec:Results}

\subsection{Mode excitation}
\label{sec:QNMoverview}

\begin{figure}
    \centering
    \includegraphics[scale=0.32]{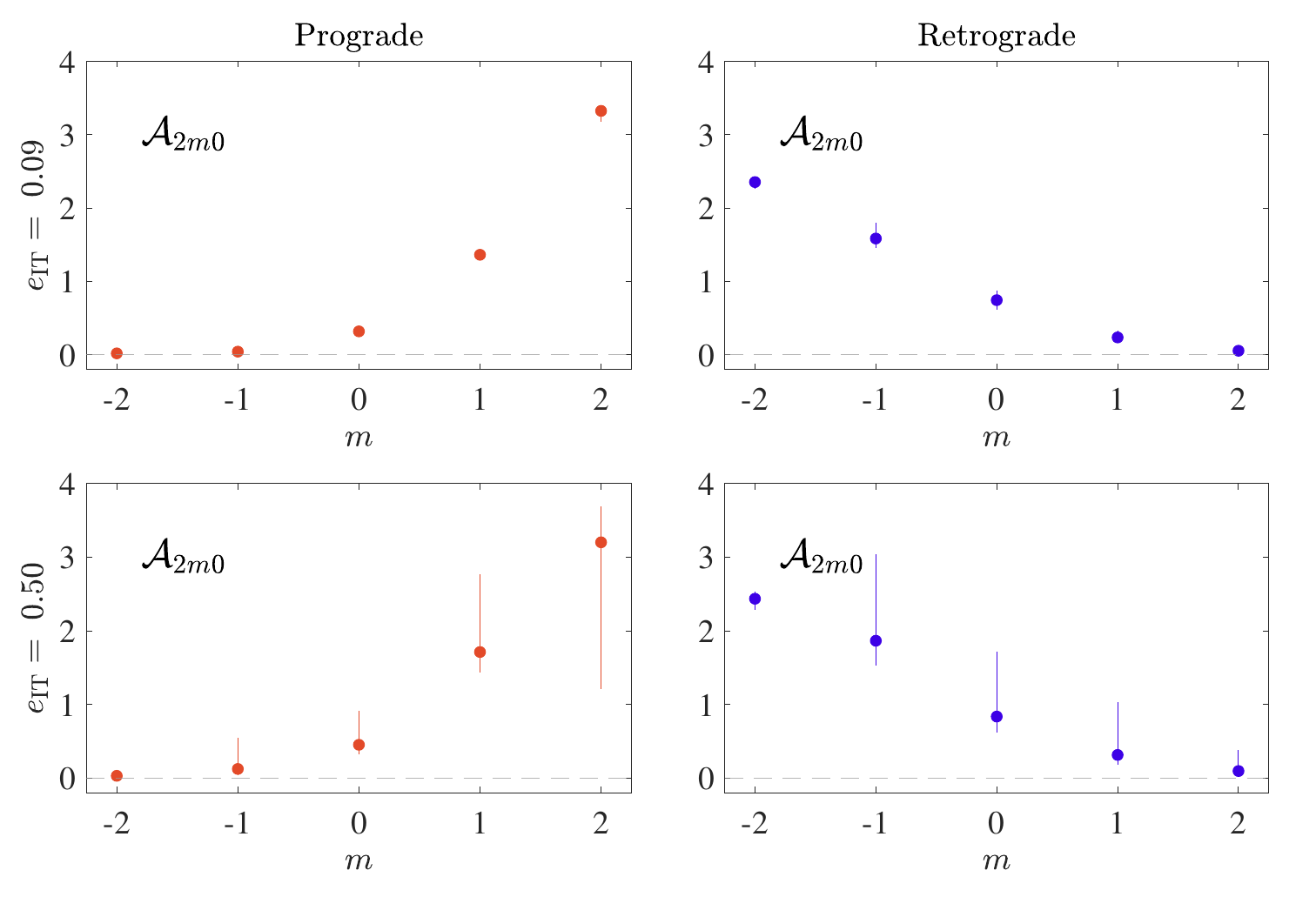}
    \caption{Prograde (left panels) and retrograde (right panels) spheroidal mode excitations for plunges at $\eta = 10^{-4}$ into a black hole with $a = 0.3M$. All data are for $k=2$ and $n = 0$ modes, with $ m\in \{-2, -1, 0, 1, 2\}$. Top row corresponds to systems with $e_{\rm IT} = 0.09$; bottom is for $e_{\rm IT} = 0.50$. The data points denote the average mode excitation across a sample of 36 trajectories that differ only by $\chi_{r\rm IT}$, and the vertical bars encode the spread in the amplitude from variations in $\chi_{r\rm IT}$ (the top of the bar gives the maximum amplitude and the bottom of the bar gives the minimum).}
    \label{fig:PGRG}
\end{figure}

Figure \ref{fig:PGRG} catalogs spheroidal QNM excitation for $k = 2$, $m\in \{-2, -1, 0, 1, 2\}$ for a system with mass ratio $\eta = 10^{-4}$ and primary spin $a = 0.3M$.  We show both prograde (left) and retrograde (right) plunges.  The top panels include results for systems with ``small'' eccentricity at plunge $e_{\rm IT} = 0.09$; bottom panels have the moderately large value $e_{\rm IT} = 0.50$.  Because an eccentric inspiral's late-time dynamics strongly depend on the value of the radial anomaly angle as the secondary enters the strong field, we calculate the mode amplitudes for a range of trajectories. Each panel shows data for the QNM excitation of 36 systems with parameters as described above, and with the anomaly angle at the end of adiabatic inspiral, $\chi_{r\rm IT} \in [0^\circ, 360^\circ]$.  Each data point denotes the sample's average mode amplitude; the vertical bars show the spread about this average.

Notice that in the small eccentricity case, the spread in mode excitation due to anomaly angle is negligible; the results are in fact very similar to mode excitation for circular and equatorial plunge (compare to the left-hand panel of Fig.\ 1 of Ref.\ \cite{Hughes2019}).  At this eccentricity, the strongest $k=2$ mode for a prograde plunge is $m=2$; for retrograde configuration, the strongest mode is $m = -2$.

For larger eccentricity, this hierarchy of mode excitation is much less clear.  Examining our results for systems with $e_{\rm IT} = 0.50$, we see that the mean amplitudes indicate that $m = 2$ dominates for prograde systems, and $m = -2$ dominates for retrograde.  The range indicated by the vertical bars show that there is a range of of $\chi_{r\rm IT}$ such that $m = 1$ dominates for prograde, and for which $m = -1$ dominates for retrograde.  In the next section, we study this dependence on $\chi_{r\rm IT}$ in more detail, exploring a range of $a$ and $e_{\rm IT}$.

\subsection{Mode excitation as a function of late-inspiral kinematics}
\label{sec:QNMresults}

\begin{figure*}
    \centering
    \includegraphics[width=\textwidth]{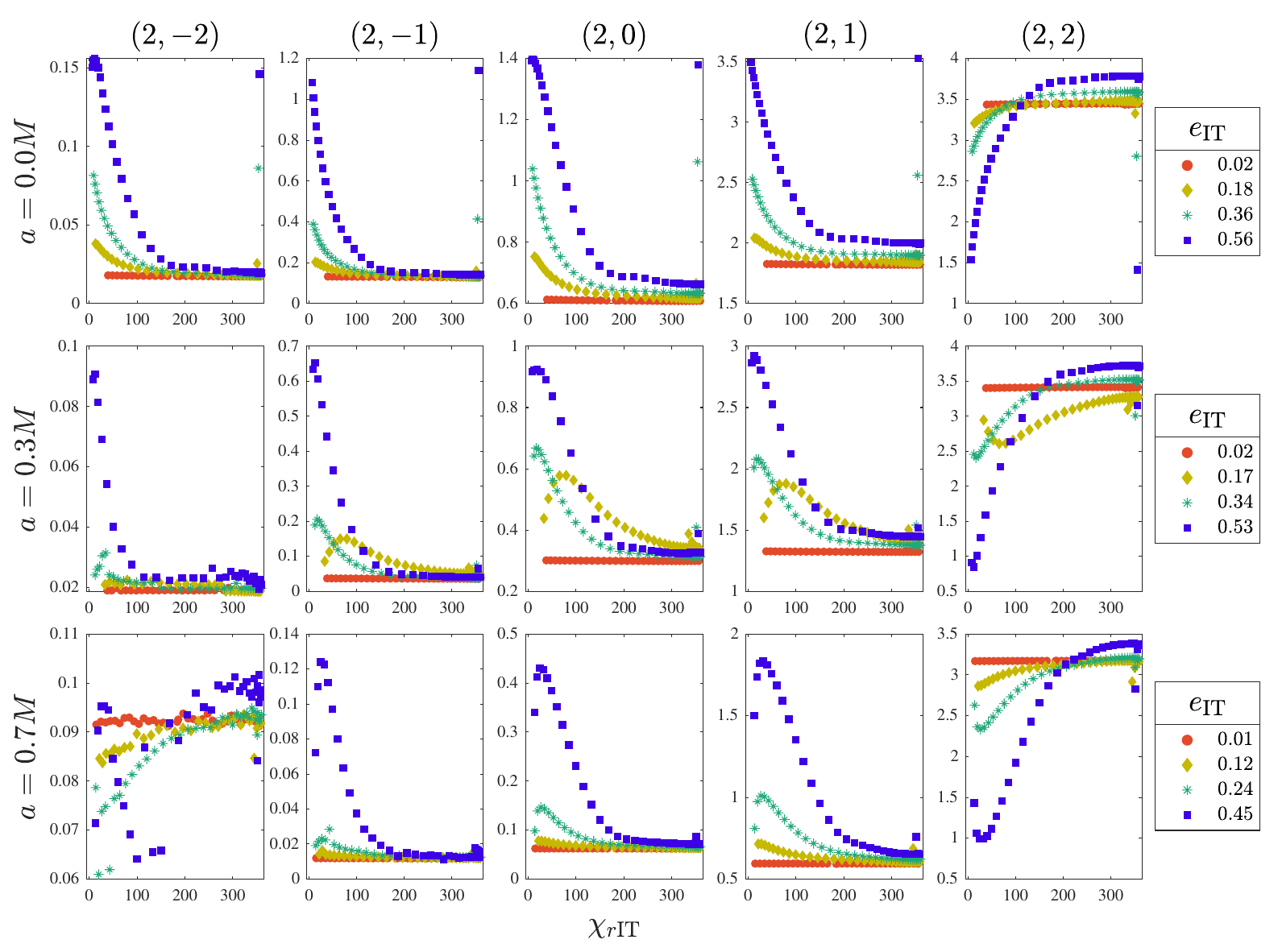}
    \caption{Mode excitation magnitude $\mathcal{A}_{km0}$ for spheroidal QNMs with $k = 2, \;m\in \{-2, -1, 0, 1, 2\}$ and $n=0$ as a function of the radial anomaly angle at the end of adiabatic inspiral, $\chi_{r\rm IT}$. Eccentricities in the strong field vary on the range $e_{\rm IT} \in [0.02, 0.56]$, with red dots denoting the lowest eccentricities, blue squares denoting the highest, and yellow diamonds and green asterisks denoting intermediate values.  Top panels show results for black hole spin $a = 0$, middle for $a = 0.3M$ and bottom for $a = 0.7M$. From left to right, $m$ varies from $-2$ to $2$. Note that the scale of the vertical axes differ between panels. In all cases, the spread in the mode amplitudes increases with eccentricity due to the dependence on $\chi_{r\rm IT}$.}
    \label{fig:3spins_chirIT_l2}
\end{figure*}

\subsubsection{Radial anomaly angle}
\label{sec:chiITqnm}

Figure \ref{fig:3spins_chirIT_l2} shows spheroidal modes with $k = 2$ for black hole spins $a\in \{0, 0.3M, 0.7M\}$.  Each case describes a prograde configuration at mass ratio $\eta = 10^{-4}$ and shows the fundamental mode amplitude $\mathcal{A}_{2m0}$ as a function of $\chi_{r\rm IT}$. From left to right, the panels include data for $m = -2$ through $m = 2$. Each panel displays QNM amplitudes from systems with four different eccentricities at the end of adiabatic inspiral in the interval $e_{\rm IT}\in [0.02, 0.56]$ (see the plot legends for precise $e_{\rm IT}$ values).  Note that the scale of the vertical axes is different in the panels; this allows us to show the more detailed structure of each mode, though it must be kept in mind that some modes are much higher amplitude than others. Appendix\ \ref{app:higherordermodes} extends this catalog to $k = 3$ QNMs.

Note the spread in mode amplitude introduced by eccentricity. Red data points show excitation for very small eccentricity and are essentially identical to those found for quasi-circular inspiral. As eccentricity increases, the mode excitation can deviate significantly from this limit.  The data are particularly clear for Schwarzschild (top row): the $(2,2)$ mode is weakest when $\chi_{r\rm IT} = 0^\circ$, increasing with this angle until the amplitude plateaus near the quasi-circular value --- though changing very suddenly as $\chi_{r\rm IT} \to 360^\circ$, enforcing continuity at $\chi_{r\rm IT} = 0^\circ$.  (We discuss this sharp change of behavior in more detail below.) By contrast, modes with $m \ne 2$ are large at $\chi_{r\rm IT} = 0^\circ$.  As this angle grows, these mode amplitudes decrease until they plateau close to the circular limit (modulo wrapping around for continuity as $\chi_{r\rm IT} \to 360^\circ$).  In all cases, either the $(2,2)$ or $(2,1)$ mode is dominant, though which one dominates varies with $\chi_{r\rm IT}$.

As we increase the primary's spin, we continue to see that eccentricity modulates QNM amplitudes depending on the value of $\chi_{r\rm IT}$, though the detailed form is comparatively less ``clean'' than in the Schwarzschild limit.  For $a = 0.7M$ (bottom row of Fig.\ \ref{fig:3spins_chirIT_l2}), the modes oscillate somewhat as $\chi_{r\rm IT}$ increases.  The $(2,2)$ mode shows moderately complicated behavior, passing through local extrema as anomaly increases.  The general trend, however, is that in eccentric systems, QNM excitation depends in a non-trivial manner on both the eccentricity at the end of inspiral and on the system's anomaly angle at this moment.  For some configurations, mode excitation is similar to the circular limit, with $(2,2)$ dominating.  For some anomaly angles, the $(2,1)$ mode dominates, and all modes differ substantially from those found in quasi-circular inspiral.

\begin{figure}[h]
    \centering
    \includegraphics[scale = 0.405]{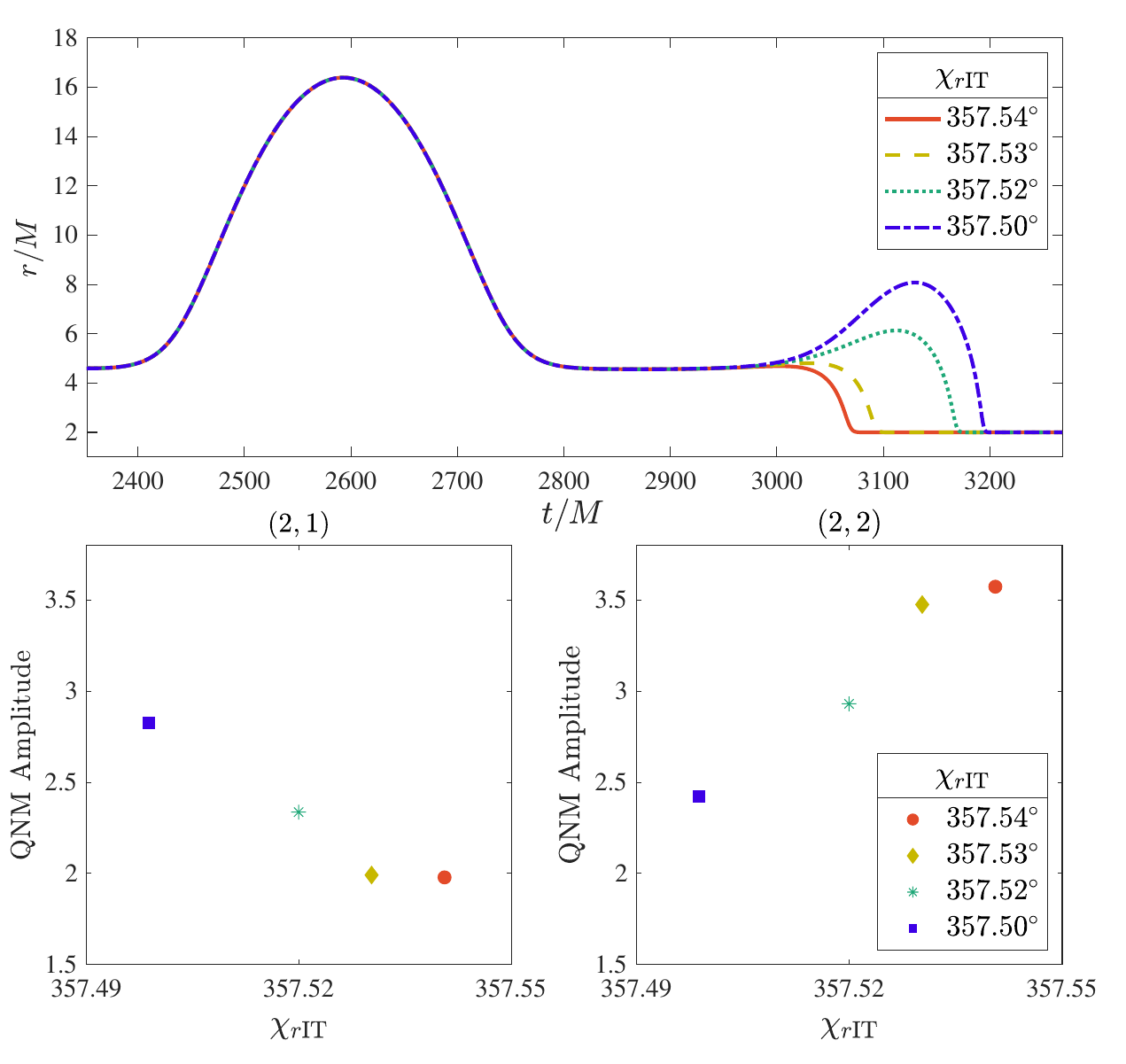}
    \caption{Worldlines and their resulting mode excitation magnitudes for spheroidal QNMs with $k = 2$, and $m = 1$ or $2$. Top panel includes trajectories for inspirals and plunges into a Schwarzschild black hole with $e_{\rm IT} = 0.56$ and $\eta = 10^{-4}$. Each trajectory has a slightly different anomaly angle in the strong field delineated by the plot legend ($\Delta \chi_{r\rm IT} \approx (10^{-2})^{\circ}$). Bottom panels show the amplitudes of the $(2,1)$ (left) and $(2,2)$ mode (right).}
    \label{fig:chicliff}
\end{figure}

For every spin, and most noticeably for $a = 0$, there is a sharp transition in mode amplitudes near $\chi_{r\rm IT} = 360^{\circ}$.  This highlights a region of parameter space where the ringdown is particularly sensitive to the anomaly angle.  To capture this behavior in more detail, Fig.\ \ref{fig:chicliff} zooms in, sampling $\chi_{r\rm IT}$ more finely.  The top panel of Fig.\ \ref{fig:chicliff} shows four worldlines for systems with $a = 0$,  $e_{\rm IT} = 0.56$ and $\eta = 10^{-4}$.  These trajectories differ only in $\chi_{r\rm IT}$, ranging from $\chi_{r\rm IT} = 357.50^\circ$ to $357.54^\circ$, stepping by $\Delta\chi_{r\rm IT} \approx 0.01^\circ$.  Notice that the time of the final plunge changes by more than $100M$ as the anomaly angle varies over this range.  The bottom panels show the amplitudes of the $(2,1)$ (left) and $(2,2)$ modes (right) for each system.  The systems' terminal kinematics and QNM excitation vary quite significantly as the systems' properties are shifted by a tiny amount.

The worldlines in Fig.\ \ref{fig:chicliff} reveal a bifurcation in $\chi_{r\rm IT}$ that shapes the late-time dynamics.  This ``chifurcation'' separates two qualitatively distinct kinematic behaviors: systems whose final kinematics have a quasi-circular character before plunge (cf.\ the solid red curve in Fig.\ \ref{fig:chicliff}, which sits at $r \simeq 4.5M$ for a time interval of several hundred $M$ before plunging), and systems whose final kinematics take the form of a nearly radial infall from relatively large radius (cf.\ the dot-dashed blue curve in this figure).  These distinct behaviors are reminiscent of critical behavior near the unstable circular orbit (UCO) explored in \cite{Faggioli2025, Gundlach2012}. In particular, the bifurcation appears to be homoclinic \cite{Levin2000, Levin2009}, as every trajectory seemingly orbits the UCO, then either plunges immediately or completes an additional radial cycle before plunging.  Appendix A of Ref.\ \cite{Faggioli2025} includes a detailed description of homoclinic orbits of Kerr black holes, highlighting the dependence of these orbits on the relativistic anomaly angle at the LSO. In both our work and Ref.\ \cite{Faggioli2025}, the chifurcation for a nonspinning system appears to occur near the last periapsis passage, which is consistent with the character of periapsis, since this is where the secondary switches from moving radially toward the primary to moving away from it. We can also schematically understand this approximate chifurcation location with an effective potential description of the motion (see Ref.\ \cite{Faggioli2025} for a more detailed discussion of strong-field black hole orbits in effective potentials and their sensitivity to the anomaly angle). In Appendix \ref{app:chifurcation}, we investigate this behavior in detail with an exploration of how sharply the ``chifurcation'' depends on mass ratio, and a brief study of the influence of spin and eccentricity at plunge on the location of the chifurcation.

Figure \ref{fig:chicliff} also clarifies an important link between late-time orbital dynamics and relative QNM excitation.  Across all spins we have studied, we consistently find that trajectories similar to the solid red trajectory, in which the final kinematics are quasi-circular and followed by a plunge from small radius, tend to be dominated by the $(2,2)$ QNM.  The relative mode excitation in such cases closely resembles the QNM spectrum found in quasi-circular coalescences --- consistent with the fact that the final several hundred $M$ of these systems' dynamics are similar to the final orbits of a quasi-circular inspiral.

In contrast, we consistently find that trajectories similar to the blue dot-dashed trajectory, in which the system's final moments resemble a direct plunge from relatively large radius with no quasi-circular ``whirling,'' can result in $(2,1)$ QNM dominance.  In the next section, we make this qualitative observation quantitative by studying the relative mode excitation as a function of the secondary's radial velocity at the PLR crossing.

\subsubsection{Mode excitation as a function of the radial velocity at the prograde light ring}
\label{sec:drdtauITqnm}

\begin{figure*}
    \centering
    \includegraphics[width=\textwidth]{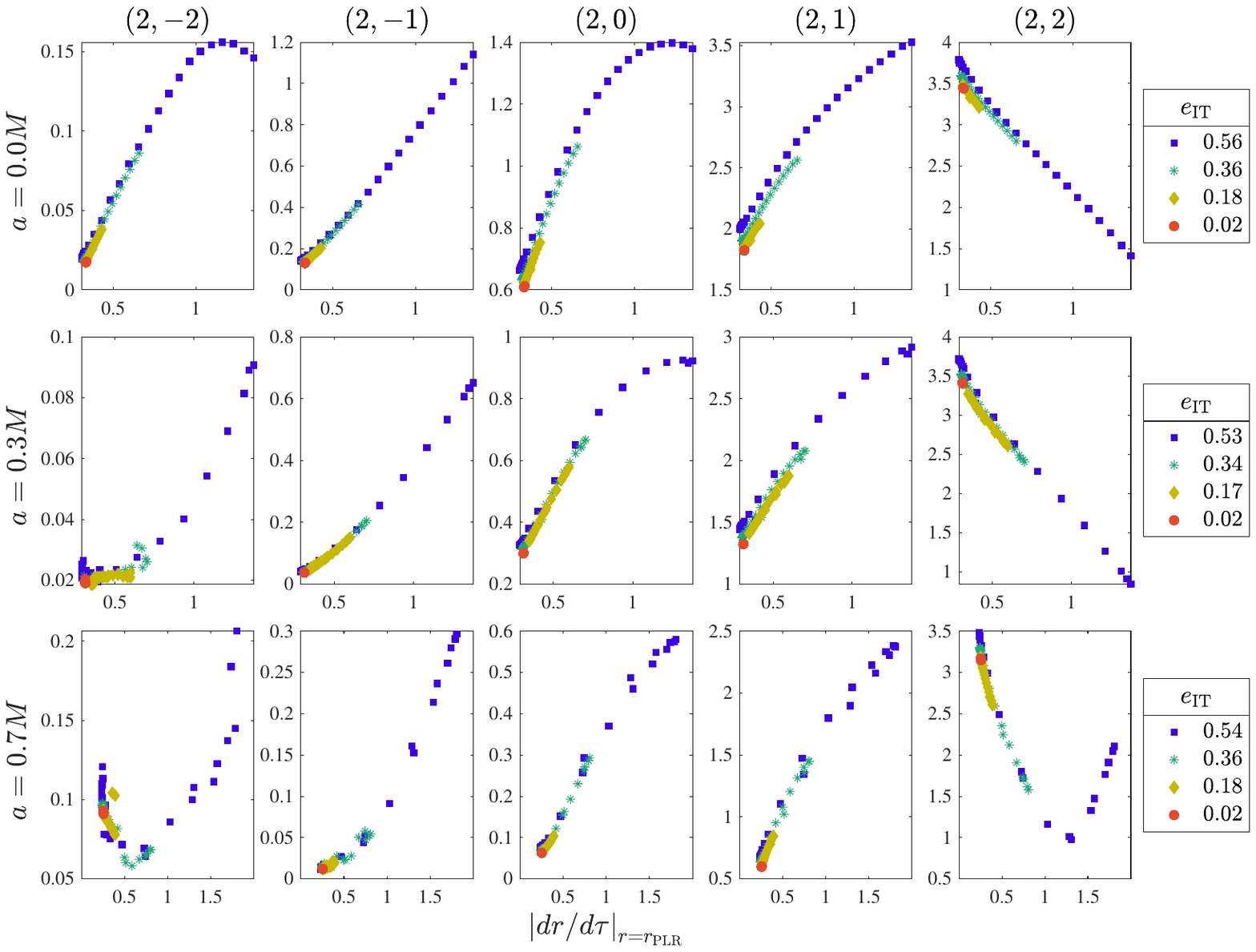}
    \caption{Mode excitation magnitude $\mathcal{A}_{km0}$ for spheroidal QNMs with $k = 2, \;m\in \{-2, -1, 0, 1, 2\}$, $n=0$ as a function of the radial velocity at the prograde equatorial light ring, $|dr/d\tau|_{r = r_{\rm PLR}}$. Eccentricities in the strong field vary on the range $e_{\rm IT} \in [0.02, 0.56]$, with red dots denoting the lowest eccentricities, blue squares the highest, and yellow diamonds and green asterisks denoting intermediate values. Top panels show results for black hole spin $a = 0$, middle panels $a = 0.3M$ and bottom panels $a = 0.7M$. From left to right, $m$ varies from $-2$ to $2$. Note that the scale of the vertical axes differs between panels.}
    \label{fig:3spins_drdtau_l2}
\end{figure*}

Figure \ref{fig:3spins_drdtau_l2} shows the same QNM excitation as is shown in Fig.\ \ref{fig:3spins_chirIT_l2}, but now presents mode amplitudes as a function of the secondary's radial velocity at the PLR $|dr/d\tau|_{r = r_{\rm PLR}}$; see Eq.\ (\ref{eqn:drdtau}) for a precise definition of this velocity and Refs.\ \cite{Khanna2017, Price2016} for background on the light ring and its relationship with QNMs.  As in Fig.\ \ref{fig:3spins_chirIT_l2}, we show the fundamental spheroidal modes with $k=2$ and $m \in \{-2, -1, 0, 1, 2\}$.  Results for $k = 3$ modes are included in Appendix \ref{app:higherordermodes}.

Begin with the $a = 0$ systems (top row of Fig.\ \ref{fig:3spins_drdtau_l2}).  The amplitude of the $(2,2)$ mode monotonically declines as $|dr/d\tau|_{r = r_{\rm PLR}}$ grows; all other modes grow with $|dr/d\tau|_{r = r_{\rm PLR}}$.  We see the same pattern in systems with $a = 0.3M$: as the radial plunge velocity increases, the $(2,2)$ mode weakens while the other $k = 2$ mode amplitudes grow with $|dr/d\tau|_{r = r_{\rm PLR}}$.  As shown in Appendix \ref{app:higherordermodes}, we tend to find analogous behavior among the $k=3$ QNMs: the amplitude of the $(3,3)$ mode falls for high plunge velocities, while other $k=3$ modes grow.

Increasing spin to $a = 0.7M$ introduces new behavior, at least for large eccentricity.  Systems with $e_{\rm IT}\in \{0.02, 0.18, 0.36\}$ follow the trends described above: the $(2,2)$ amplitudes decrease as $|dr/d\tau|_{r = r_{\rm PLR}}$ increases, while the other $m$ modes grow.  However, for the $e_{\rm IT} = 0.54$ case, we see that the $(2,2)$ mode decreases until $|dr/d\tau|_{r = r_{\rm PLR}} \simeq 1.3$, at which point its evolution turns around and it increases as $|dr/d\tau|_{r = r_{\rm PLR}}$ grows.  (The other $k = 2$ modes evolve in a fairly monotonic fashion, mostly increasing with radial plunge velocity.)  It is worth noting that these cases correspond to the highest plunge velocities found in this sample.  It appears that as the plunge velocity gets very large, these systems excite more complicated QNM spectra.

\subsection{Tail Excitation}
\label{sec:tailresults}

\begin{figure}
    \centering
    \includegraphics[scale = 0.435]{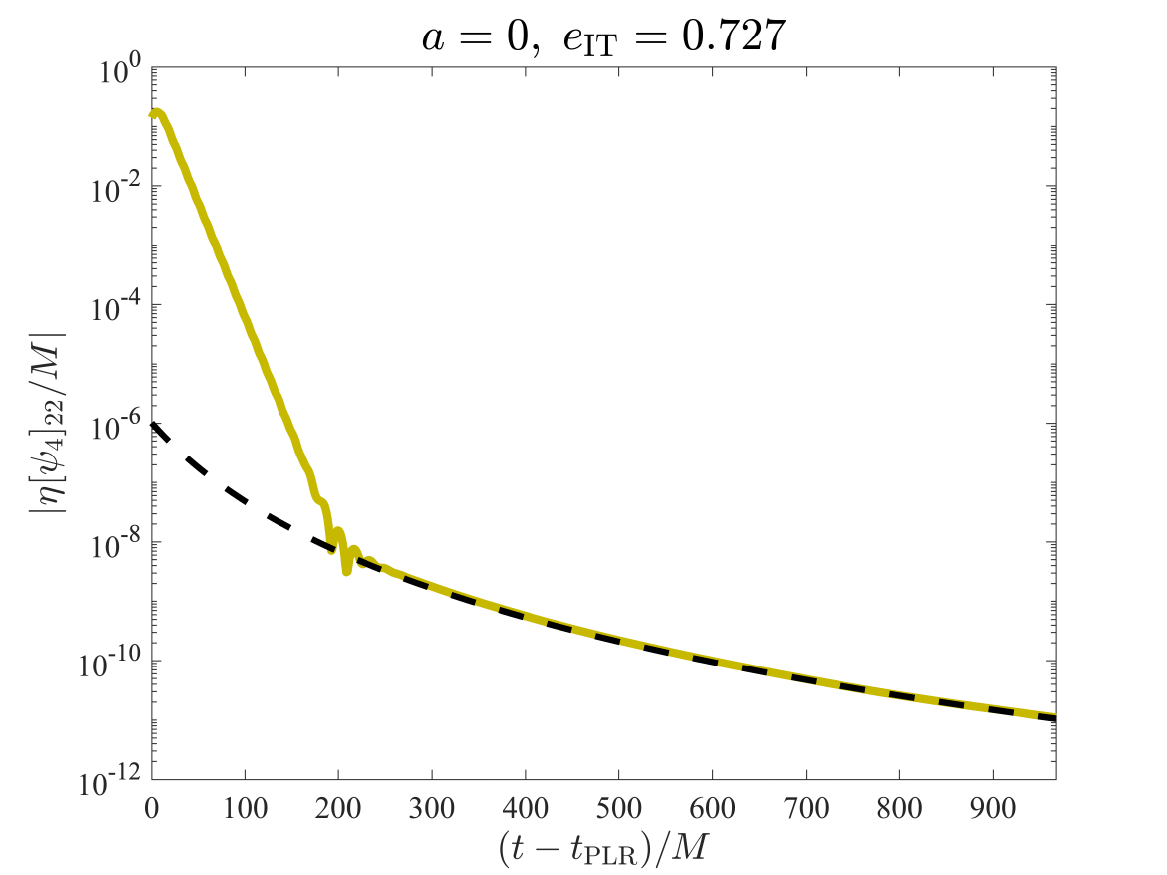}
    \caption{Ringdown amplitude from a merger with $a = 0$, $\eta = 10^{-4}$, $e_{\rm IT} = 0.727$ and $\chi_{r\rm IT} = 89^{\circ}$. The solid yellow line traces the amplitude of $[\psi_4]_{22}$ as a function of time past the PLR crossing. The black dashed curve represents a fit of the late-time tail to a model with $\mathcal{A}_{\psi \rm tail}^{22} = 3.28\times10^{5}\pm 2.7\times 10^3$, $c_{\rm tail}^{22} = 132 \pm 0.11$ and $p_{\psi \rm tail }^{22} = - 5.42 \pm 1.1\times 10^{-3}$.}
    \label{fig:tailfit}
\end{figure}

We begin our study of the late ringdown by presenting a typical Price tail from our dataset and its power-law fit.  Figure \ref{fig:tailfit} shows the post-merger amplitude $[\psi_4]_{22}$ sourced by a small body merging with a Schwarzschild black hole at eccentricity $e_{\rm IT} = 0.727$ and mass ratio $\eta = 10^{-4}$. The secondary ends its adiabatic inspiral at $\chi_{r\rm IT} = 89^{\circ}$; In Sec.\ \ref{sec:chiITtails}, we explore how this system's tail varies with $\chi_{r\rm IT}$.  Focus first on the mode's amplitude, the solid yellow line.  For $t \in [t_{\rm PLR}, t_{\rm PLR}+150M]$, the amplitude rapidly drops as QNMs decay. After this point, but before $t\approx t_{\rm PLR} + 300M$, the ringdown is in an oscillatory regime where QNM and tail contributions mix. We exclude this region, as well as the preceding QNM-dominated region, from our tail fit. Once $t \gtrsim t_{\rm PLR} + 300M$, the waveform shows a clear power-law decay as the system settles into equilibrium. We model this portion of the ringdown with Eq.\ (\ref{eqn:psi4tails}); the dashed black line in Fig.\ \ref{fig:tailfit} shows the fit we find. On the interval $t-t_{\rm PLR} \in [300 M, 750M]$, we find $\mathcal{A}_{\psi \rm tail}^{22} = 3.28\times10^{5}\pm 2.7\times 10^3$, $c_{\rm tail}^{22} = 132 \pm 0.11$ and $p_{\psi \rm tail }^{22} = - 5.42 \pm 1.1\times 10^{-3}$. When we start the fit at later times, the power-law index increases toward its expected value of $-6$.  We do not undertake a detailed analysis with different fitting intervals here; see IFK25 for such an analysis, specifically with respect to the asymptotic behavior of the power law.

\subsubsection{Tails as a function of eccentricity and spin}
\label{TailsResultsEcc}

\begin{figure*}[t]
    \includegraphics[scale = 0.455]{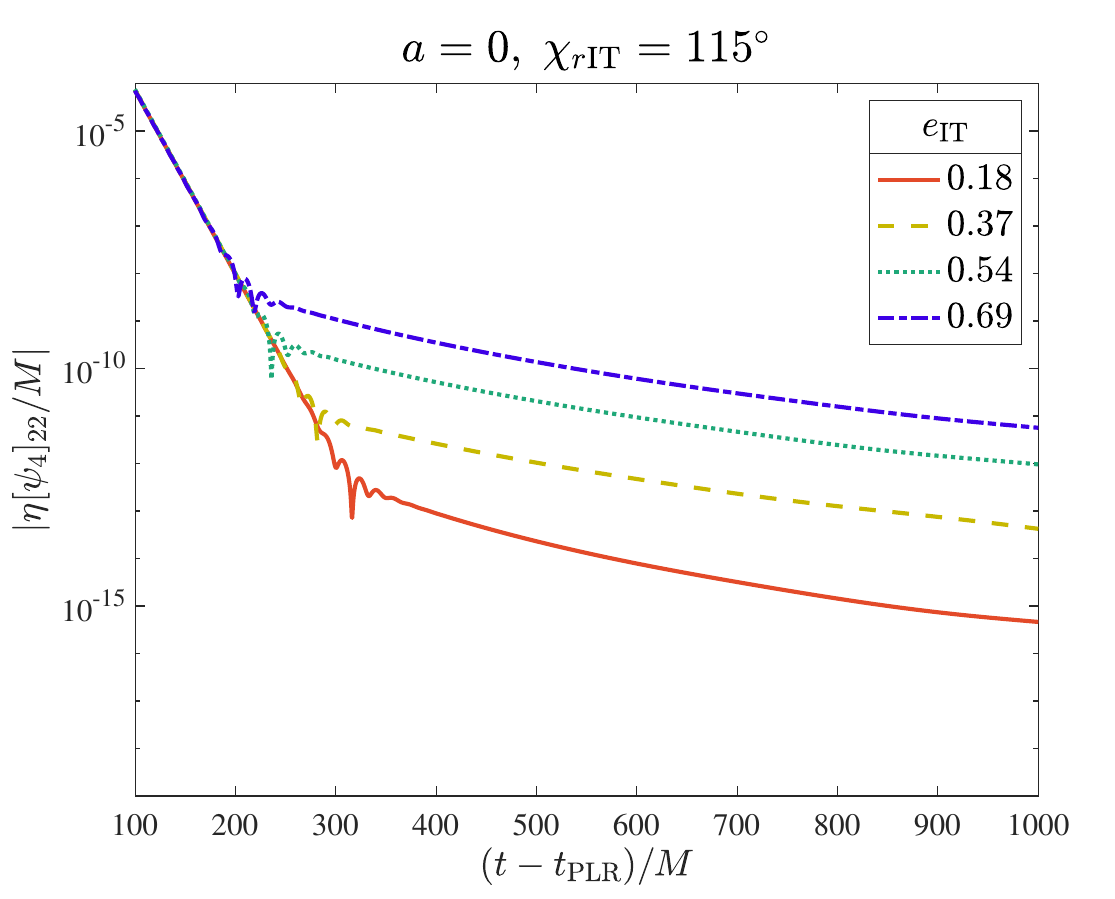}
    \hfill 
    \includegraphics[scale = 0.465]{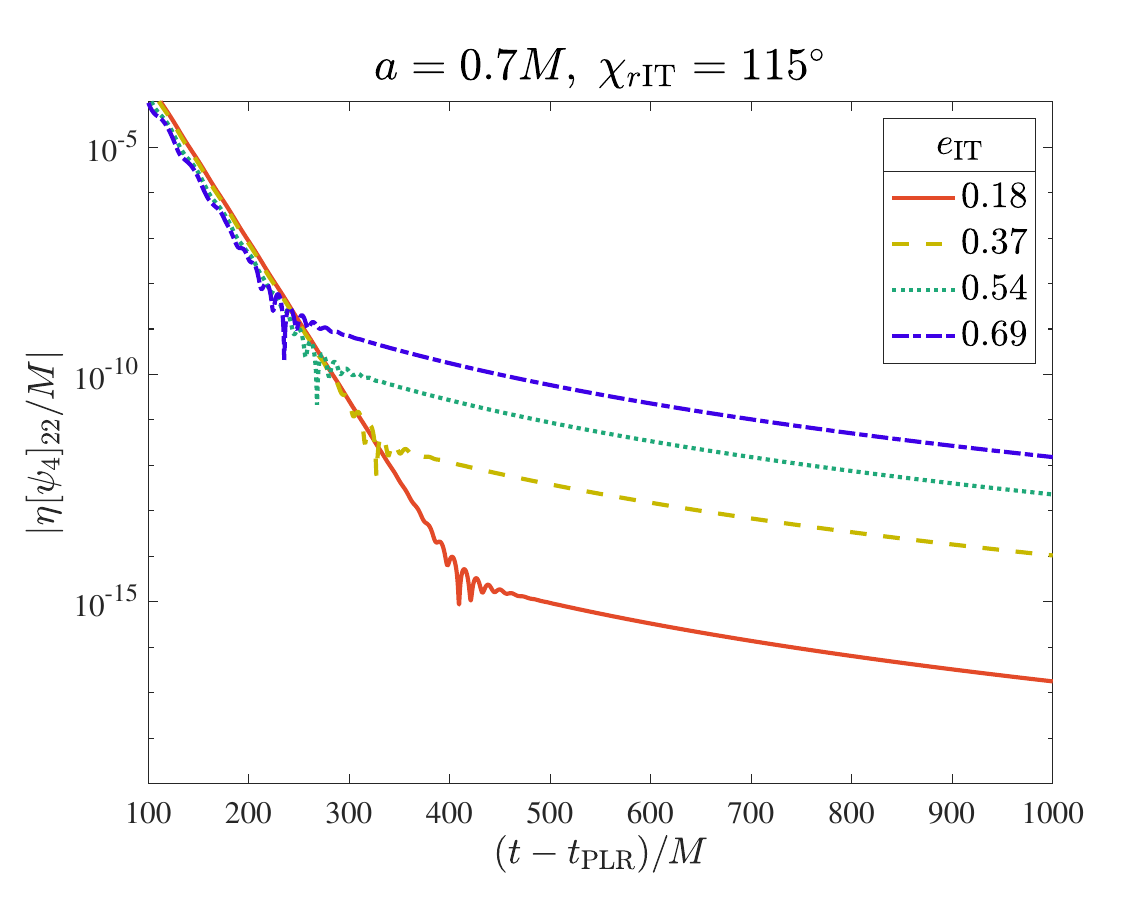}
    \caption{Power-law tails as a function of eccentricity at the end of adiabatic inspiral.  In each case, $\chi_{r\rm IT} = 115^{\circ}$ and $\eta = 10^{-4}$. Left panel shows results for secondary's plunging into a Schwarzschild black hole; Right panel shows results for secondary's plunging into a Kerr black hole with $a = 0.7M$.}
\label{fig:tails_eITcompare}
    \end{figure*}

Figure \ref{fig:tails_eITcompare} shows the ringdown amplitude after the PLR crossing for binaries in a variety of orbits, all with $\eta = 10^{-4}$ and $\chi_{r\rm IT} = 115^{\circ}$. Each system in the left-hand panel has $a = 0$; each curve presents a different eccentricity from the set $e_{\rm IT} \in \{0.18, 0.37, 0.54, 0.69\}$. The right-hand panel shows results from systems with $a = 0.7M$, but are otherwise identical to those on the left.

We fit each of these tails to a power law; Table \ref{table:tails_eITcompare} gives the values of $\mathcal{A}_{\psi \rm tail}$, $c_{\rm tail}$, and $p_{\psi \rm tail }$ we find fitting on the interval $t-t_{\rm PLR}\in [t_{\rm{start}}, 1000M]$, where $t_{\rm{start}}$ is tabulated for each system. For both spins, there is a clear correlation between eccentricity and tail amplitude, at least at fixed radial anomaly: larger eccentricity means larger tail amplitude.  This eccentricity-induced tail amplification is in complete agreement with Refs.\ \cite{Islam2024, DeAmicis2024A, DeAmicis2024B,Albanesi2023} and IFK25.  Also in line with these studies, we find that the tail dominates the ringdown earlier for systems with higher eccentricities.

The final fit parameter we consider, $p_{\psi \rm tail }$, does not vary significantly as eccentricity changes, except for the Schwarzschild $e_{\rm IT} = 0.18$ ringdown, where the slope associated with the power law is slightly more shallow than in our other systems.  The deviations we find could simply result from our choice of fitting interval for each system, since the differences we observe are of a similar order of magnitude to those expected from fitting interval shifts; see discussion in Ref.\ \cite{Islam2024} and in IFK25. Regarding the tail's dependence on spin, our results indicate that while spin does affect power-law tails, the specific relationship is not straightforward. For $e_{\rm IT} = 0.54$ and $e_{\rm IT} = 0.69$, the tails for the $a = 0$ and $a = 0.7M$ systems are nearly indistinguishable despite notable differences in the QNM and oscillatory regions. Note that IFK25 investigates systems with higher spins than we display here, finding particularly strong tails for retrograde captures into black holes with $a = 0.9M$.

\begin{table*}
\begin{center}
\renewcommand{\arraystretch}{1.3}
\begin{tabular}{||c c c c c c||}
 \hline
 $a/M$ & $e_{\rm IT}$ & $\mathcal{A}_{\psi\rm tail}$ & $c_{\rm tail}$ & $|p_{\psi \rm{tail}}|$ & $t_{\rm start}/M$ \\ [0.5ex] 
 \Xhline{4\arrayrulewidth} 
 0 & 0.18 \quad & $3.68\pm 0.08$ \quad & $-45.9 \pm 0.28 $ \quad & $5.36 \pm 3.1\times 10^{-3}$ \quad & 434  \\ 
 \hline
 0 & 0.37 \quad & $1.55 \times 10^4 \pm 574$ \quad & $79.4 \pm 0.60$ \quad & $5.72 \pm 4.9\times 10^{-3}$ \quad & 404 \\
 \hline
  0 & 0.54 \quad & $8.50 \times 10^4 \pm 1.9\times 10^{3}$ \quad & $161 \pm 0.33$ \quad & $5.55 \pm 3.0\times 10^{-3}$ \quad & 340 \\
 \hline
 0 & 0.69 \quad & $2.48 \times 10^6 \pm 1.7\times 10^{4}$ \quad & $169 \pm 0.09$ \quad & $5.76 \pm 9.2\times 10^{-4}$ \quad & 319 \\
 \hline
 0.7 & 0.18 \quad & $7.45 \pm 0.04$ \quad & $22.8 \pm 0.08$ \quad & $5.86 \pm 6.8\times 10^{-4}$ \quad & 602 \\
 \hline
  0.7 & 0.37 \quad & $935 \pm 0.88$ \quad & $61.1 \pm 0.02$ \quad & $5.61 \pm 1.3\times 10^{-4}$ \quad & 502 \\
 \hline
  0.7 & 0.54 \quad & $1.69 \times 10^4 \pm 12$ \quad & $50.4 \pm 0.01$ \quad & $5.59 \pm 9.7\times 10^{-5}$ \quad & 413 \\
 \hline
  0.7 & 0.69 \quad & $1.18 \times 10^6 \pm 1.5\times 10^{3}$ \quad & $-173 \pm 0.02$ \quad & $5.89 \pm 1.7\times 10^{-4}$ \quad & 401 \\
 \hline
\end{tabular}
\end{center}
\caption{Power-law tail fit parameters for the systems displayed in Fig.\ \ref{fig:tails_eITcompare}.}
\label{table:tails_eITcompare}
\end{table*}

\subsubsection{Tails as a function of the radial anomaly angle}
\label{sec:chiITtails}

\begin{figure*}
    \centering
    \includegraphics[width=\textwidth]{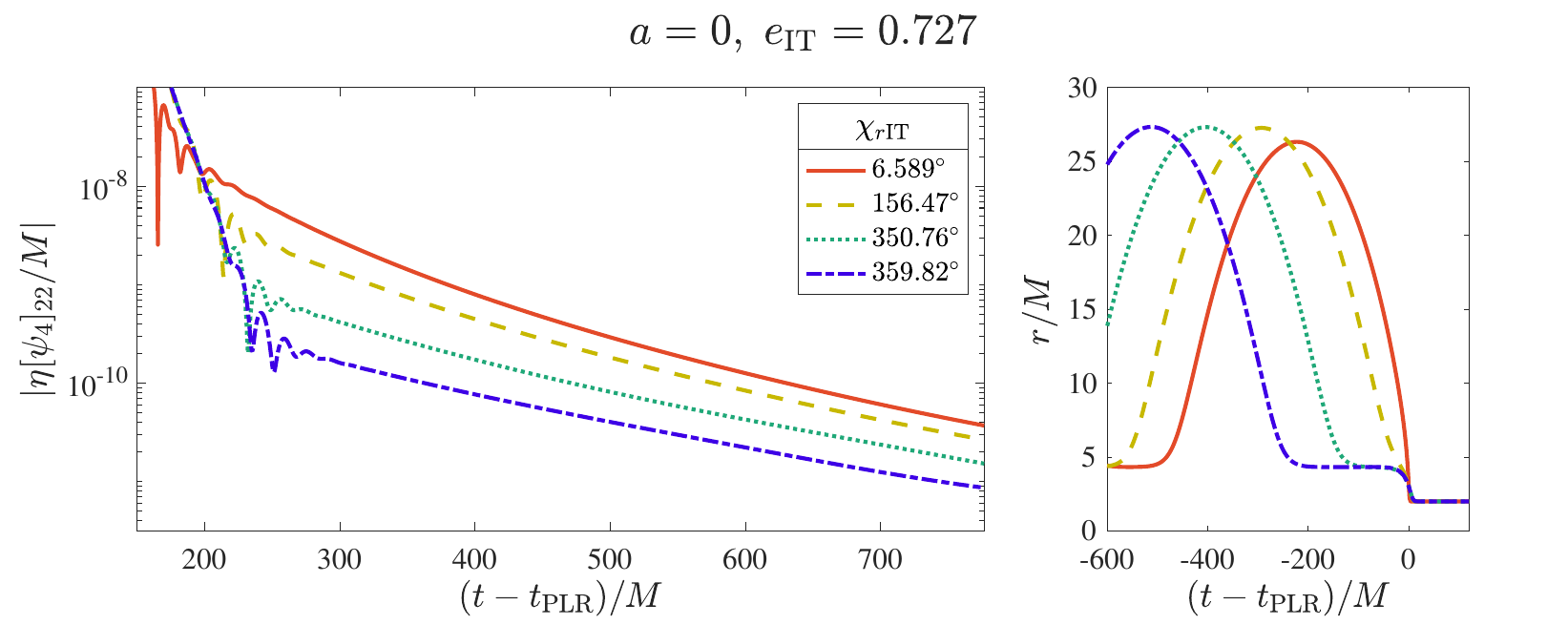}
    \caption{Left panel: $[\psi_4]_{22}$ ringdown amplitude for four nonspinning binaries, all with $e_{\rm IT} = 0.727$ and $\eta = 10^{-4}$, but different values of the radial anomaly angle in the strong field. Right panel: the worldlines associated with the ringdown amplitudes in the left panel.}
    \label{fig:tails_chi}
\end{figure*}

We next examine power-law tails from a set of eccentric binaries, holding all parameters fixed except the late-time radial anomaly angle $\chi_{r\rm IT}$.  Both panels of Fig.\  \ref{fig:tails_chi}, describe systems with $a = 0$, $e_{\rm IT} = 0.727$, and $\eta = 10^{-4}$.  The left-hand panel shows the post-merger amplitude $[\psi_4]_{22}$ as a function of time after the PLR crossing sourced by the worldlines shown in the right-hand panel.  These binaries are distinct both in their late-time behavior and their excited ringdown amplitudes, despite the fact that they only differ by $\chi_{r\rm IT}$.  Table \ref{table:tails_chiITcompare} gives the tail parameters we find for these systems.  Visual inspection of the tails paired with the fit parameters we find shows that a binary's terminal dynamics strongly shapes the morphology of its power-law tail.

Systems that end with a radial plunge from a large radius (the solid red and dashed yellow trajectories) excite stronger tails than systems that complete quasi-circular whirls before plunging (the dotted green curve and dot-dashed blue curve). Moreover, there appears to be a relationship between the time spent executing quasi-circular whirls and the amplitude of the excited tail, as the system with the longest quasi-circular whirl excites the tail with the lowest amplitude (note that $\mathcal{A}_{\psi\textrm{tail}}$ is inversely related to the amplitude of the tail when it decouples from the oscillatory regime).

The variations in $\mathcal{A}_{\psi \rm tail}$ and $c_{\rm tail}$ that arise from changes in $\chi_{r\rm IT}$ emphasize that neither eccentricity $e_{\rm IT}$ nor radial anomaly $\chi_{r\rm IT}$ is enough to determine a system's tail: the impact of these parameters on the tail is strongly correlated.  Notice that the four tails shown here asymptotically decay with very similar slopes, though a more thorough analysis extending deeper into the ringdown is needed to ascertain whether they asymptote to the expected value $p_{\psi\rm tail} = -6$.  We outline plans for a more detailed study of the relationship between the tail morphology and the plunge dynamics in our conclusions.

\begin{table*}
\begin{center}
\renewcommand{\arraystretch}{1.3}
\begin{tabular}{||c c c c c||}
 \hline
 $\chi_{r\rm IT}$ [deg]  & $\mathcal{A}_{\psi\rm tail}$ & $c_{\rm tail}$ & $|p_{\psi \rm{tail}}|$ & $t_{\rm start}/M$ \\ [0.5ex] 
  \Xhline{4\arrayrulewidth} 
6.589 & $1.44\times 10^{5}\pm 1.2\times 10^{3}$ \quad & $90.0 \pm 0.12$ \quad & $5.31 \pm 1.1\times 10^{-3}$ & 400  \\ 
 \hline
156.47 & $1.81 \times 10^6 \pm 6.2\times 10^{4}$ \quad & $187 \pm 0.58$ \quad & $5.65 \pm 4.5\times 10^{-3}$ & 425 \\
 \hline
350.76 & $4.27 \times 10^8 \pm 7.0\times 10^{7}$ \quad & $422 \pm 3.2$ \quad & $6.32 \pm 2.1\times 10^{-2}$ & 435 \\
 \hline
359.82 & $1.42 \times 10^{10} \pm 5.7\times 10^{9}$ \quad & $591 \pm 8.6$ \quad & $6.77 \pm 5.0\times 10^{-2}$ & 445 \\

 \hline
\end{tabular}
\end{center}
\caption{Power-law tail fit parameters for the systems displayed in Fig.\ \ref{fig:tails_chi}.}
\label{table:tails_chiITcompare}
\end{table*}

\section{Conclusions and Next Steps}
\label{sec:Conclusion}

In this analysis, we have investigated how orbital eccentricity and the radial anomaly angle affect late-time waveforms from binary black hole coalescences in the small-mass-ratio limit.  We construct inspiral-transition-plunge worldlines with the BH25 procedure for eccentric small-mass-ratio binaries, then generate the gravitational waveforms produced by these worldlines using a time-domain Teukolsky equation solver.  We focus on the final ringdown cycles of these waves, seeking to understand whether eccentricity leaves a distinguishable imprint on the excitation of Kerr QNMs and the late-time Price tails.  We find that eccentricity can influence QNM amplitudes and the structure of the tail, but the details of this influence are complicated by dependence on the strong-field radial anomaly angle.  This result is not surprising: our previous work studying eccentric dynamics, BH25, found that the kinematics of the late inspiral and plunge of eccentric systems depend strongly on the anomaly.  Some eccentric binaries merge after a final cycle which terminates in a plunge from large radius; others plunge from small radius following some number of quasi-circular whirls. The radial anomaly angle at which the secondary enters the strong field controls which of these possibilities the secondary undertakes, which in turn determines the nature of the system's ringdown.

Secondaries that plunge immediately after completing quasi-circular whirls excite QNM spectra that closely mimic the mode structure of purely quasi-circular mergers.  In particular, the $(2,2)$ mode dominates.  These systems can have significant eccentricity during the inspiral, but if inspiral ends at an anomaly angle such that the system whirls at periapse before plunging, the ringdown appears quasi-circular.  If the same secondary approaches the strong field with a different anomaly, it may instead plunge from large radius after a final radial cycle.  In this case, the spectrum is dominated by the $(2,1)$ QNM.  In addition to depending on anomaly angle, these behaviors depend on the system's spin.  High spins make $(2,1)$ dominance possible at lower eccentricities.

In a similar vein, we find that the power-law tails from systems that plunge from large radius have larger amplitudes and earlier start times than systems that execute quasi-circular whirls right before plunge.  Though eccentricity can amplify the power-law tail, as found by many related works \cite{Islam2024, DeAmicis2024A, DeAmicis2024B, Albanesi2023}, the anomaly angle also impacts the tail morphology.  This indicates that the eccentricity-based amplification is fundamentally determined by a binary's detailed terminal dynamics, rather than simply the value of its eccentricity.

A key parameter controlling an eccentric coalescence's ringdown appears to be the velocity of the final plunge.  This velocity is in turn influenced by the primary's spin, orbital eccentricity, and the late-time anomaly.  We find a clear relationship between the plunge's radial velocity through the prograde light ring and the excitation of $k = 2$ QNMs: the amplitude of the $(2,2)$ mode decreases as this velocity increases, and all other $k = 2$ modes grow with plunge velocity.  Systems that merge after quasi-circular whirls typically result in plunges with lower radial velocities, and are dominated by the $(2,2)$ QNM.  Systems that plunge from large radius after a final radial oscillation cross the light ring with quite large radial velocity, and the $(2,1)$ QNM tends to dominate.  As other studies of eccentric waveforms using alternate frameworks and other mass ratios (cf.\ Refs.\ \cite{Faggioli2025, Nee2025, Carullo2024qnms}) examine the relative excitation of QNMs, we look forward to cross-checking results and investigating how well these results hold up as a function of mass ratio.

A natural direction for future work is to combine this analysis with that of Refs.\ \cite{Lim2019, ApteHughes2019} and BH25 to investigate the relative excitation of QNMs from systems with generic orbital configurations --- systems that are both eccentric and inclined.  In Sec.\ VII of BH25, we outline our plan to use \cite{ApteHughes2019} to update our eccentric transition-to-plunge procedure to describe inclined systems, thereby allowing the production of generic inspiral-transition-plunge worldlines.  Calculating these generic worldlines requires generic radiation reaction data grids, which are now under development for LISA mission data studies (e.g.\ \cite{Chapmanbird2025}).  Once generic worldlines exist, it will not be difficult to construct their associated waveforms with the Teukolsky equation, and use the algorithm of \cite{Lim2019} to study the relative QNM excitation.  We are particularly interested in assessing potential degeneracies between inclination and eccentricity in the QNM hierarchy.  When studied independently, it appears that inclination can leave a stronger imprint on the QNM spectrum than eccentricity.  A comprehensive study that includes both of these orbital parameters could verify if this is indeed the case, as well as help to uncover what precisely can be learned about a binary's coalescence geometry from the ringdown alone.

Another key area of interest across mass ratios is the QNM excitation from binaries where both black holes are spinning, especially in cases with spin precession. The self-force community has made excellent progress in recent years in calculating inspiral waveforms from spinning bodies orbiting Kerr black holes (see, for example, Refs.\ \cite{Drummond2024, Piovano2024, Matthews2025, Skoupy2025}).  As these studies finalize the details of the inspiral, we can begin working towards a description of the transition to plunge of a spinning body, and eventually apply such a description to a ringdown study.

A more detailed understanding from the perturbative limit of late-time spin precession may be a useful tool for informing NR studies, which have shown that highly precessing systems can excite the $(2,\pm 1)$ and $(2,0)$ fundamental QNMs \cite{Zhu2025, OShaughnessy2013, Nobili2025}. Since eccentricity and inclination can also add power to these specific QNMs, a fully generic study that includes spin precession would of course be ideal.  This work could be useful for informing analyses of high-mass GW events such as GW190521 \cite{GW190521} and GW231123 \cite{GW231123}, as the GW signals from both of these events may show evidence of spin precession. Studies of these systems thus far have been limited by the small number of inspiral cycles in the LVK band, increasing the importance of the ringdown in uncovering the progenitors of high-mass LVK binaries. 

It may also be valuable to undertake a systematic study of how the power-law tails depend on, and presumably encode, the binary's late-time kinematics.  The tail-fitting algorithm developed in IFK25 could be used to examine how the combined dependence on radial anomaly angle and eccentricity influences the behavior of the parameters which describe the tail.  Because we find that the tail depends most strongly on the final inspiral and plunge kinematics, it is plausible that a clean dependence on the radial velocity during the plunge exists as well.




\section*{Acknowledgments}

We thank Halston Lim for assisting with our implementation of the quasinormal mode extraction pipeline described in Ref.\ \cite{Lim2019}. D.R.B.\ and S.A.H.\ acknowledge support from NSF Grants No. PHY--2110384 and PHY--2409644. G.K.\ acknowledges support from NSF Grants No.\ PHY--2307236 and DMS--2309609.  We also thank Tousif Islam and Guglielmo Faggioli for sharing a draft of Ref.\ \cite{Islam2025} with us, and engaging in much helpful discussion.  Computations used GNU Parallel \cite{Tange2011a}, MIT Kavli Institute resources at MIT's {\tt engaging} computing cluster, {\tt subMIT} resources at MIT Physics, and the UMass-URI {\tt UNITY} HPC/AI cluster at the Massachusetts Green High Performance Computing Center (MGHPCC).  

\appendix

\section{Higher Order Mode Excitation}
\label{app:higherordermodes}

\begin{figure*}[ht]
    \centering
    \includegraphics[width=\textwidth]{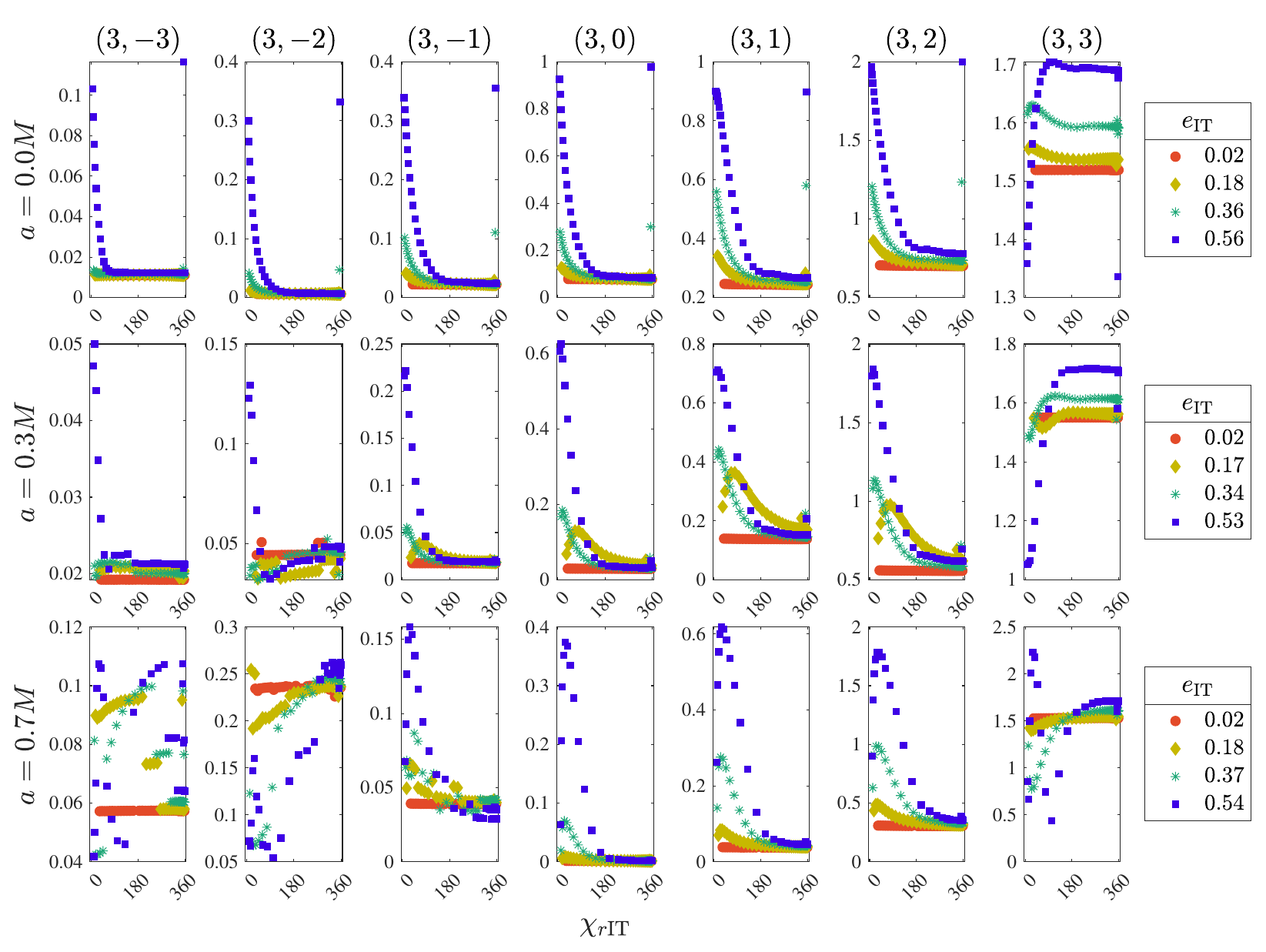}
    \caption{Mode excitation magnitude $\mathcal{A}_{km0}$ for spheroidal QNMs with $k=3, \;m\in \{-3, -2, -1, 0, 1, 2, 3\}$, $n=0$ as a function of radial anomaly at the end of adiabatic inspiral, $\chi_{r\rm IT}$. Eccentricities in the strong field vary on the range $e_{\rm IT} \in [0.01, 0.56]$.  Red dots denote the lowest eccentricities, blue squares the highest, and yellow diamonds and green asterisks denote intermediate values.  Top panels show results for black hole spin $a = 0$, middle panels $a = 0.3M$, and bottom panels $a = 0.7M$. From left to right, $m$ varies from $-3$ to $3$.  Note that the scale of the vertical axes differs between panels.}
    \label{fig:3spins_chirIT_l3}
\end{figure*}

\begin{figure*}
    \centering
    \includegraphics[width=\textwidth]{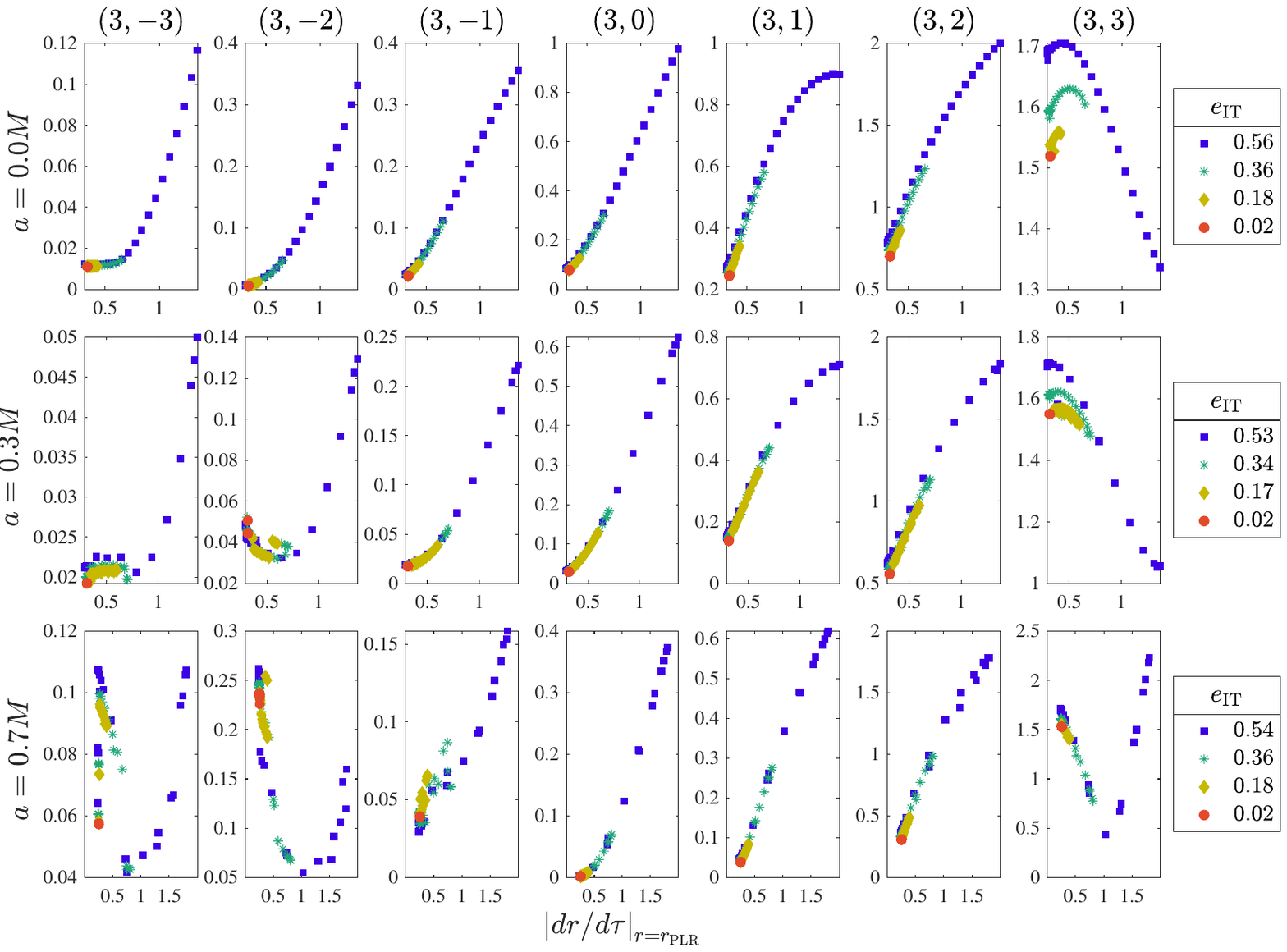}
    \caption{Mode excitation magnitude $\mathcal{A}_{km0}$ for spheroidal QNMs with $k=3, \;m\in \{-3, -2, -1, 0, 1, 2, 3\}$, $n=0$ as a function of radial velocity at the prograde equatorial light ring, $|dr/d\tau|_{r = r_{\rm PLR}}$.  Eccentricities in the strong field vary on the range $e_{\rm IT} \in [0.01, 0.56]$.  Red dots denote the lowest eccentricities, blue squares the highest, and yellow diamonds and green asterisks denote intermediate values.  Top panels show results for black hole spin $a = 0$, middle panels $a = 0.3M$, and bottom panels $a = 0.7M$. From left to right, $m$ varies from $-3$ to $3$.  Note that the scale of the vertical axes differs between panels.}
    \label{fig:3spins_drdtau_l3}
\end{figure*}

In this appendix, we catalog higher-order mode excitation for the systems explored in Sec.\ \ref{sec:QNMresults}.  Figure\ \ref{fig:3spins_chirIT_l3} shows the relative QNM amplitudes for fundamental spheroidal modes with $k = 3$ and $m\in \{-3, ..., 3\}$ as a function of the radial anomaly angle at the end of adiabatic inspiral. Every system has $\eta = 10^{-4}$ and $e_{\rm IT}\in [0.02, 0.56]$. Top panels focus on binaries with $a = 0$, middle panels have $a = 0.3M$, and bottom panels have $a = 0.7M$.  As was the case with the $k = 2$ spheroidal modes, Fig.\ \ref{fig:3spins_chirIT_l3} showcases the spread in the absolute QNM amplitudes sourced by variations in $\chi_{r\rm IT}$. 

To connect the relative QNM excitation to the binary's coalescence dynamics, Fig.\ \ref{fig:3spins_drdtau_l3} shows the same spheroidal mode amplitudes from Fig.\ \ref{fig:3spins_chirIT_l3} as a function of the secondary's radial velocity at the prograde light ring, $|dr/d\tau|_{r = r_{\rm PLR}}$. Focusing first on the nonspinning results in the top panels, the relationship between the QNM amplitudes and the plunge velocity is clear: larger radial velocities through the light ring add power to every mode but $(k,m) = (3, 3)$, which dominates the $k = 3$ excitation for quasi-circular mergers.  This pattern echoes what we observed in the $k=2$ mode excitation: the dominant $k=2$ mode in the quasi-circular case, the $(2,2)$ mode, is less excited when the secondary rapidly passes through the light ring, while all other $k=2$ modes gain power.

Combining our $k=2$ and $k=3$ results allows us to take a more detailed inventory of eccentric QNM excitation. In nonspinning systems whose final orbital cycles have the character of quasi-circular whirls and in which the systems cross the light ring with relatively low plunge velocity, the $(2,2)$ QNM dominates, followed by the $(2,1)$ mode and then $(3,3)$.  Systems that execute a final radial cycle then plunge from large radius tend to dominantly excite $(2,1)$, then $(2,2)$, followed by $(3,2)$ and $(3,3)$; it is worth noting that we only observe this QNM hierarchy when $e_{\rm IT} \gtrsim 0.5$ for $a=0$.  Increasing the spin of the primary to $a = 0.7M$ does not alter the leading mode, but the ordering of the subdominant modes can shift, with the $(3,3)$ mode surpassing the $(2,1)$ when the final kinematics before plunge are quasi-circular.  The detailed dependence of both the $(2,2)$ and $(3,3)$ modes on $|dr/d\tau|_{r = r_{\rm PLR}}$ increases in complexity for higher spins and plunge velocities, making the subdominant QNM hierarchy for systems that plunge from large radius less straightforward.  Also note that for $a = 0.7M$, eccentricities as low as $e_{\rm IT} \approx 0.35$ can produce a QNM spectrum that is discernible from the quasi-circular case if the radial anomaly allows us to distinguish the late-time coalescence kinematics.

\section{``Chifurcation''}
\label{app:chifurcation}

In Sec.\ \ref{sec:Results}, we explored the excitation of QNMs and late-time tails as a function of the radial anomaly at the end of adiabatic inspiral, $\chi_{r\rm IT}$.  Ringdowns from eccentric systems that differ only by $\chi_{r\rm IT}$ can present distinct QNM amplitudes and tail morphologies.  Further, in Sec.\ \ref{sec:chiITqnm} we found a region of parameter space in which the worldline is especially sensitive to $\chi_{r\rm IT}$, resulting in a bifurcation in the late-time kinematics that is particularly consequential to the ringdown. This bifurcation with respect to $\chi_{r\rm IT}$, which we have dubbed the ``chifurcation,'' represents an extreme case of an eccentric inspiral's sensitivity to initial conditions.  In this appendix, we study the chifurcation in more detail, first focusing on its dependence on the binary's mass ratio.

\begin{figure}
    \centering
    \includegraphics[scale = 0.40]{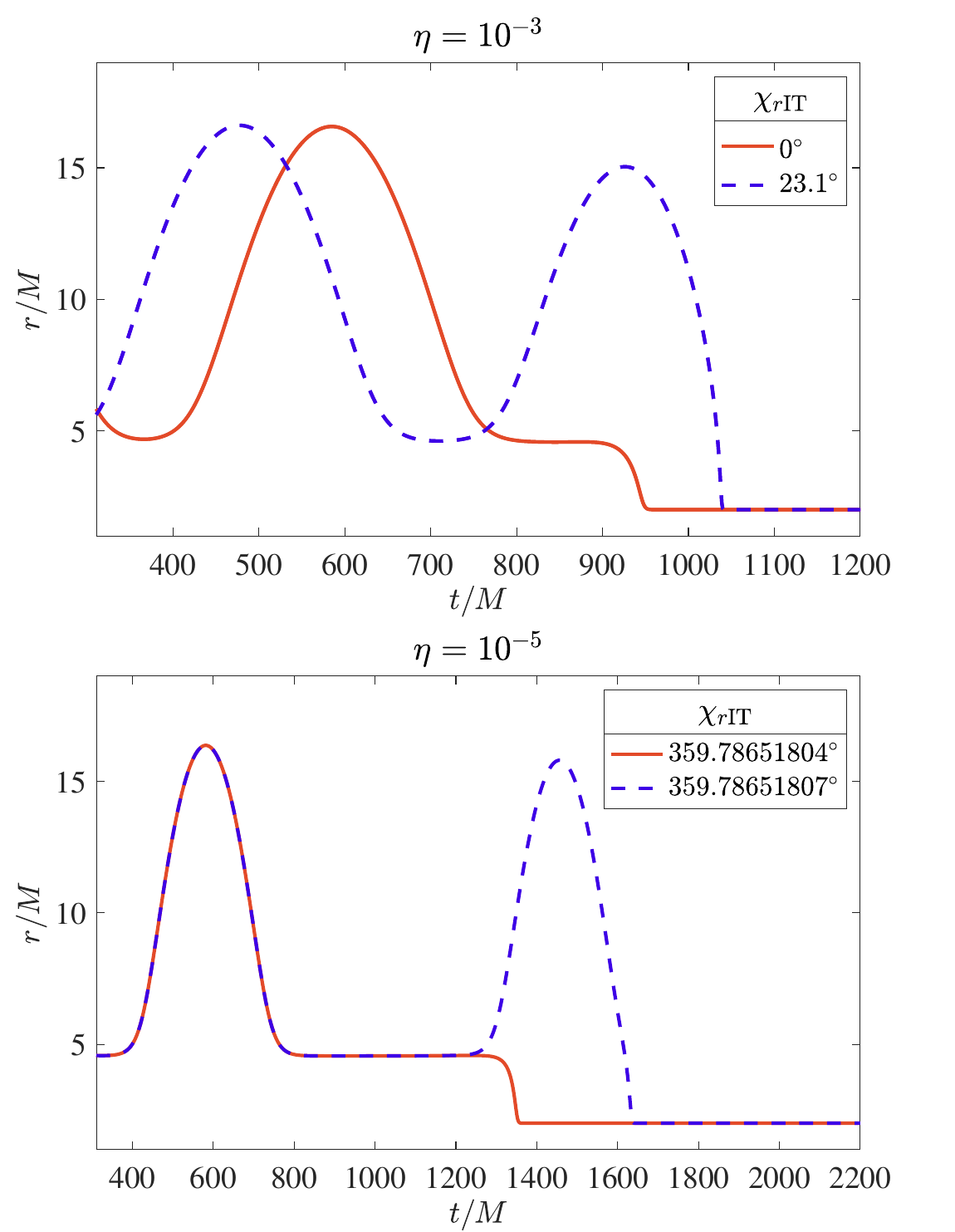}
    \caption{Worldline sensitivity to the parameter $\chi_{r\rm IT}$ as a function of mass ratio $\eta$. Every worldline corresponds to the inspiral, transition, and plunge of a small body into a Schwarzschild black hole at eccentricity $e_{\rm IT} = 0.56$. Top panel shows systems with mass ratio $\eta = 10^{-3}$; bottom shows $\eta = 10^{-5}$. }
    \label{fig:chifurcation_eta}
\end{figure}

Figure \ref{fig:chifurcation_eta} examines the relationship between the chifurcation and the mass ratio of the binary. Each worldline shown in Fig.\ \ref{fig:chifurcation_eta} describes a system with $a = 0$ and $e_{\rm IT} = 0.56$.  Those presented in the top panel have $\eta = 10^{-3}$; those in the bottom have $\eta = 10^{-5}$.  Both mass ratios exhibit the behaviors separated by the chifurcation: orbits that complete quasi-circular whirls at plunge, the solid red line in both panels, and orbits that plunge after retreating to large radii, the dashed blue curves. Trajectories with anomaly angles in between the given values either whirl slightly longer than the solid red trajectory, or execute an additional radial cycle with a smaller amplitude than the dashed blue curve (see Fig.\ \ref{fig:chicliff} for examples of this intermediate behavior). For example, if one of the systems in the top panel instead had $\chi_{r\rm IT} = 10^{\circ}$, it would execute a radial cycle to about $r = 11M$ before plunging.  For the dashed blue trajectories, we chose trajectories with a final apoapsis passage at $r \approx 15M$ for ease of comparison between mass ratios.

From this brief study, we can see that mass ratio determines how sensitive the worldlines are to the relativistic anomaly angle: for $\eta = 10^{-3}$, the two behaviors are separated by $\Delta\chi_{r\rm IT} \approx 23^{\circ}$; at the more extreme mass ratio, $\Delta\chi_{r\rm IT} \approx (10^{-8})^{\circ}$. The different degrees of sensitivity are also apparent by the dephasing in the worldlines: the two trajectories in the bottom panel are so close in $\chi_{r\rm IT}$ that up until their final orbital cycle, they are indistinguishable at the resolution of this plot; the worldlines in the top panel are out of phase throughout the inspiral.  These results, paired with the $\eta = 10^{-4}$ case with $\Delta\chi_{r\rm IT} \approx (10^{-2})^{\circ}$ of Sec.\ \ref{sec:chiITqnm}, indicate that the chifurcation is far more sensitive for EMRI systems than for binaries with more comparable masses.

The quasi-circular motion executed in Fig.\ \ref{fig:chifurcation_eta} (i.e., the $\Delta t \sim$ several hundred $M$ interval of motion at nearly constant $r$) is reminiscent of ``zoom-whirl'' orbit behavior, in which the secondary completes several quasi-circular whirls around the primary during its pericenter passage \cite{Glampedakis2002}.  As can be seen in Fig.\ \ref{fig:chifurcation_eta}, the duration of this whirling increases as the mass ratio approaches the test-mass limit.  We can estimate the number of whirls with $N_w = \Delta\phi/2\pi$, where $\Delta \phi$ is the axial angle accumulated during the segment of nearly constant $r$.  For example, for the worldlines with $\eta = 10^{-5}$ in Fig.\ \ref{fig:chifurcation_eta}, $N_w = [\phi(t = 1230M) - \phi(t = 790M)]/2\pi$.  The $\eta = 10^{-3}$ worldlines shown here have $N_w\approx 1.5$; the $\eta = 10^{-5}$ trajectories have $N_w\approx 6.8$.  The more extreme mass-ratio systems exhibit more pronounced zoom-whirl behavior in the strong field than their comparable-mass counterparts.  Indeed, the comparable mass systems presented in Ref.\ \cite{Nee2025} do not exhibit prolonged quasi-circular kinematics before plunge for any choice of orbital anomaly.  The strongly ``chifurcating'' trajectories are most pronounced in the small mass ratio limit.

We also carried out a brief investigation into the influence of a system's mass ratio, spin and late-time eccentricity on the chifurcation angle. Although we always find that the chifurcation occurs slightly before periapsis, its precise location can differ between systems with different $\eta, a$ and $e_{\rm IT}$. In Table \ref{table:chifurcation}, we show angular intervals $X^{\rm chif}_{r\rm{IT}}$ that bracket the chifurcation for a variety of systems. For a given $\eta, a$ and $e_{\rm IT}$, systems with $\chi_{r\rm{IT}}$ at the lower bound of $X^{\rm crit}_{r\rm{IT}}$ execute quasi-circular whirls before plunging, while systems with $\chi_{r\rm{IT}}$ at the upper bound of $X^{\rm crit}_{r\rm{IT}}$ move radially outward before plunging. The transition between these two behaviors, the chifurcation, thus occurs at some $\chi_{r\rm{IT}}\in X^{\rm crit}_{r\rm{IT}}$. While narrowing the width of each interval $X^{\rm chif}_{r\rm{IT}}$ is possible with repeated calculations, the intervals do not overlap for the cases we wish to compare, so the given resolution of $X^{\rm chif}_{r\rm{IT}}$ is sufficient. The $X^{\rm chif}_{r\rm{IT}}$ intervals in Table \ref{table:chifurcation} indicate that increasing the spin shifts the chifurcation to smaller $\chi_{r\rm IT}$, while increasing the eccentricity shifts the chifurcation to larger $\chi_{r\rm IT}$.  Although we only examined two mass ratios, $\eta = 10^{-4}$ and $\eta = 3\times 10^{-4}$, what we see is that, holding all other parameters constant, making the mass ratio less extreme tends to move the chifurcation to a point earlier than periapsis (i.e., to a somewhat smaller value of $\chi^{\rm chif}_{r\rm{IT}}$).

\begin{table}
\renewcommand{\arraystretch}{1.3}
\begin{tabular}{||c c c c||}
 \hline
 $\eta$ & \hspace*{1em} $a/M$ & \hspace*{1em} $e_{\rm IT}$ & \hspace*{2em}Bracketing interval $X^{\textrm{chif}}_{r\textrm{IT}}$ [deg] \\ [0.3ex] 
 
\Xhline{5\arrayrulewidth} 
 $10^{-4}$ & \hspace*{1em} 0  & \hspace*{1em} 0.27  & \hspace*{2em} $353.18 < \chi_{r\rm IT}^{\rm chif}<  353.34$\\
 \hline
  $10^{-4}$ & \hspace*{1em} 0  & \hspace*{1em} 0.46  & \hspace*{2em} $355.24 < \chi_{r\rm IT}^{\rm chif}< 355.40 $\\ 
 \hline
   $10^{-4}$ & \hspace*{1em} 0 & \hspace*{1em} 0.67  & \hspace*{2em} $ 356.27 < \chi_{r\rm IT}^{\rm chif}<  356.34$\\ 
   
\Xhline{3\arrayrulewidth}
   $10^{-4}$ & \hspace*{1em} 0.1 & \hspace*{1em} 0.27 & \hspace*{2em} $352.80 < \chi_{r\rm IT}^{\rm chif}<  352.96$\\ 
 \hline 
   $10^{-4}$ & \hspace*{1em} 0.1  & \hspace*{1em} 0.46  & \hspace*{2em} $ 354.98 < \chi_{r\rm IT}^{\rm chif}<  355.11$\\ 
 \hline
    $10^{-4}$ & \hspace*{1em} 0.1 & \hspace*{1em} 0.67  & \hspace*{2em}$ 356.08 < \chi_{r\rm IT}^{\rm chif}<  356.18$\\ 
    
\Xhline{3\arrayrulewidth}
    $10^{-4}$ & \hspace*{1em} 0.3 & \hspace*{1em} 0.27  & \hspace*{2em} $352.29 < \chi_{r\rm IT}^{\rm chif}<  352.48$\\ 
 \hline
     $10^{-4}$ & \hspace*{1em} 0.3 & \hspace*{1em} 0.46  & \hspace*{2em} $354.72 < \chi_{r\rm IT}^{\rm chif}<  354.87$\\ 
 \hline
      $10^{-4}$ & \hspace*{1em} 0.3 & \hspace*{1em} 0.67  & \hspace*{2em} $355.85 < \chi_{r\rm IT}^{\rm chif}< 355.95$\\ 
      
\Xhline{3\arrayrulewidth}
      $10^{-4}$ & \hspace*{1em} 0.7 & \hspace*{1em} 0.27  & \hspace*{2em} $ 350.71 < \chi_{r\rm IT}^{\rm chif} < 350.93$\\ 
 \hline
      $10^{-4}$ & \hspace*{1em} 0.7 & \hspace*{1em}0.46  & \hspace*{2em} $353.81 < \chi_{r\rm IT}^{\rm chif}<  353.96$\\ 
 \hline
      $10^{-4}$ & \hspace*{1em} 0.7 & \hspace*{1em} 0.67  & \hspace*{2em} $355.21 < \chi_{r\rm IT}^{\rm chif}<  355.30$\\ 
      
\Xhline{3\arrayrulewidth}
    $3\times 10^{-4}$ & \hspace*{1em} 0 & \hspace*{1em} 0.27  & \hspace*{2em} $350.32< \chi_{r\rm IT}^{\rm chif}<  350.59 $\\ 
 \hline
     $3\times 10^{-4}$ & \hspace*{1em} 0 & \hspace*{1em}0.67  & \hspace*{2em} $354.80 < \chi_{r\rm IT}^{\rm chif}< 355.01$\\ 
     
\Xhline{3\arrayrulewidth}
    $3\times 10^{-4}$ & \hspace*{1em} 0.3 & \hspace*{1em}0.27 & \hspace*{2em} $ 349.28 < \chi_{r\rm IT}^{\rm chif}< 349.55$\\ 
    \hline
    $3\times 10^{-4}$ & \hspace*{1em} 0.3 & \hspace*{1em} 0.67 & \hspace*{2em} $354.16 < \chi_{r\rm IT}^{\rm chif}< 354.37$\\ 
    
\Xhline{3\arrayrulewidth}
    $3\times 10^{-4}$ & \hspace*{1em} 0.7 & \hspace*{1em}0.27 & \hspace*{2em} $ 347.06 < \chi_{r\rm IT}^{\rm chif}< 347.36$\\ 
 \hline
     $3\times 10^{-4}$ & \hspace*{1em} 0.7 & \hspace*{1em}0.67 & \hspace*{2em} $353.19 < \chi_{r\rm IT}^{\rm chif}< 353.40$\\ 
 \hline
\end{tabular}
\caption{Angular intervals $X^{\rm chif}_{r\rm{IT}}$ that bracket the chifurcation for systems with different $\eta, a$ and $e_{\rm IT}$.}
\label{table:chifurcation}
\end{table}



\section{Mode excitation as a function of {\it ad hoc} model parameter choices}
\label{app:alphabeta}

As we stressed in BH25 and recount in Sec.\ \ref{sec:WL}, our procedure for generating worldlines relies on two \textit{ad hoc} parameters: a parameter $\alpha_{\rm IT}$ which determines when adiabatic inspiral ends and the transition begins; and a parameter $\beta_{\rm TP}$, which sets the end of the transition.  See BH25 for the detailed definitions and formulae describing these quantities.  In this appendix, we investigate the robustness of our ringdown model against variations in these model parameter choices, focusing on the QNM-dominated regime. 

\begin{figure*}
    \centering
    \includegraphics[width=\textwidth]{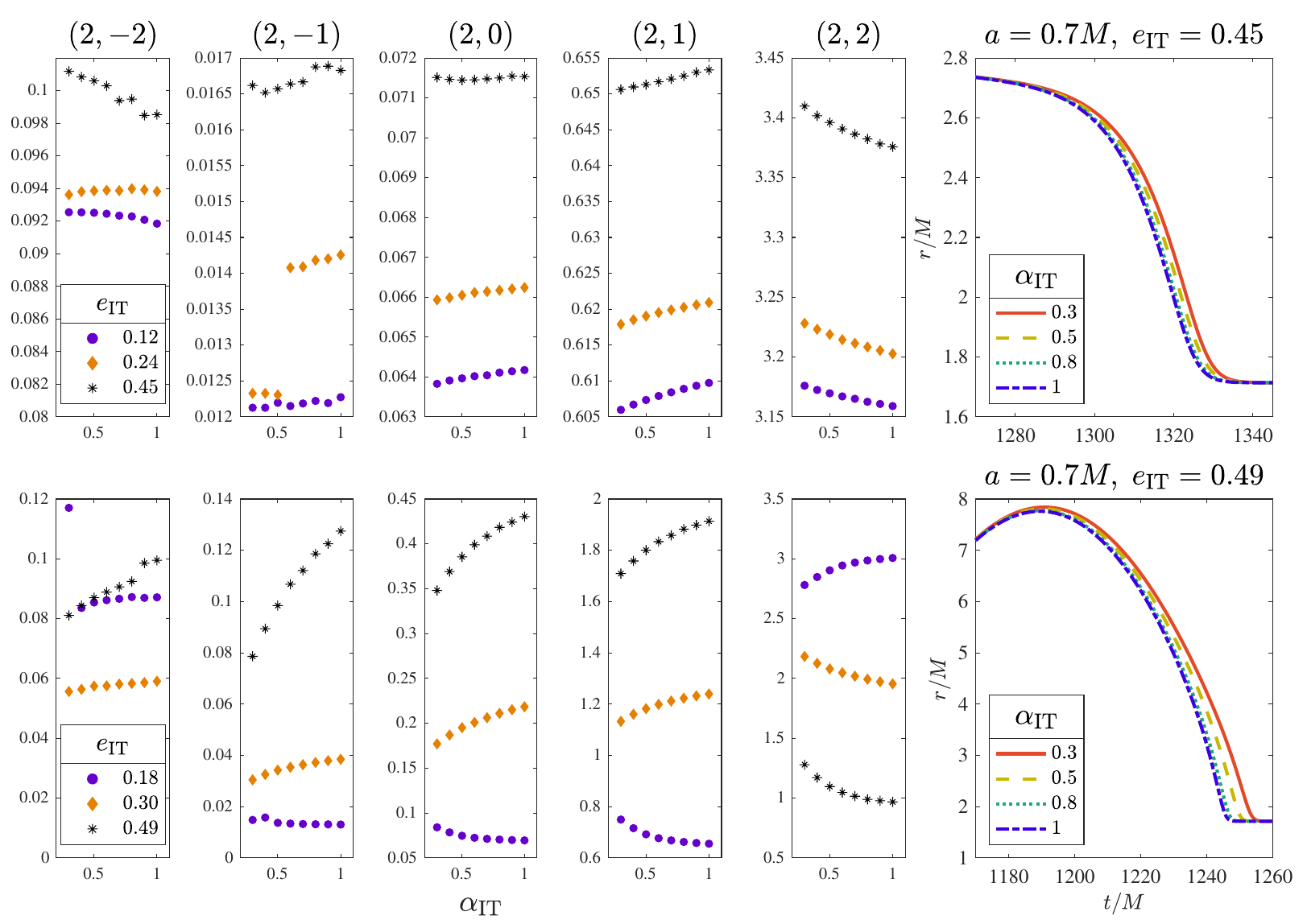}
    \caption{Mode excitation magnitude $\mathcal{A}_{km0}$ for spheroidal QNMs with $k = 2, \;m\in \{-2, -1, 0, 1, 2\}$, $n=0$ as a function of $\alpha_{\rm IT}$. All systems have $a = 0.7M$, $\eta = 10^{-4}$ and $e_{\rm IT}\in [0.12, 0.49]$.  Systems described by the top panels immediately plunge when adiabatic inspiral ends (as depicted by the worldlines in the right-most panel), and systems included in the bottom panels execute one final radial cycle before plunging (as illustrated in the right-most panel).}
    \label{fig:aIT}
\end{figure*}

Figure \ref{fig:aIT} examines how the $k = 2$ QNMs depend on $\alpha_{\rm IT}\in[0.1, 1.1]$ for a variety of eccentric mergers with $a = 0.7M$ and $\eta = 10^{-4}$.  Since the eccentric transition permits two classes of late-time dynamics, we study both scenarios in this figure: systems that complete quasi-circular whirls before the plunge are described by the top panels, and systems that complete a radial cycle before plunging from large radius are in the bottom panels.  In both cases, the left five panels show the QNM amplitudes; and the right-most panels show four representative worldlines. While variations in $\alpha_{\rm IT}$ can affect the absolute QNM amplitude, especially in the cases shown in the bottom panels, the relative mode excitation is unchanged: if the $(2,2)$ mode dominates when $\alpha_{\rm IT} = 0.1$, it remains the strongest excited mode when $\alpha_{\rm IT} = 1.1$. Bearing in mind that BH25 recommended holding $0.75 \le \alpha_{\rm IT} \le 1$, we see that the QNM excitations are quite stable against variations in $\alpha_{\rm IT}$.

\begin{figure*}
    \centering
    \includegraphics[width=\textwidth]{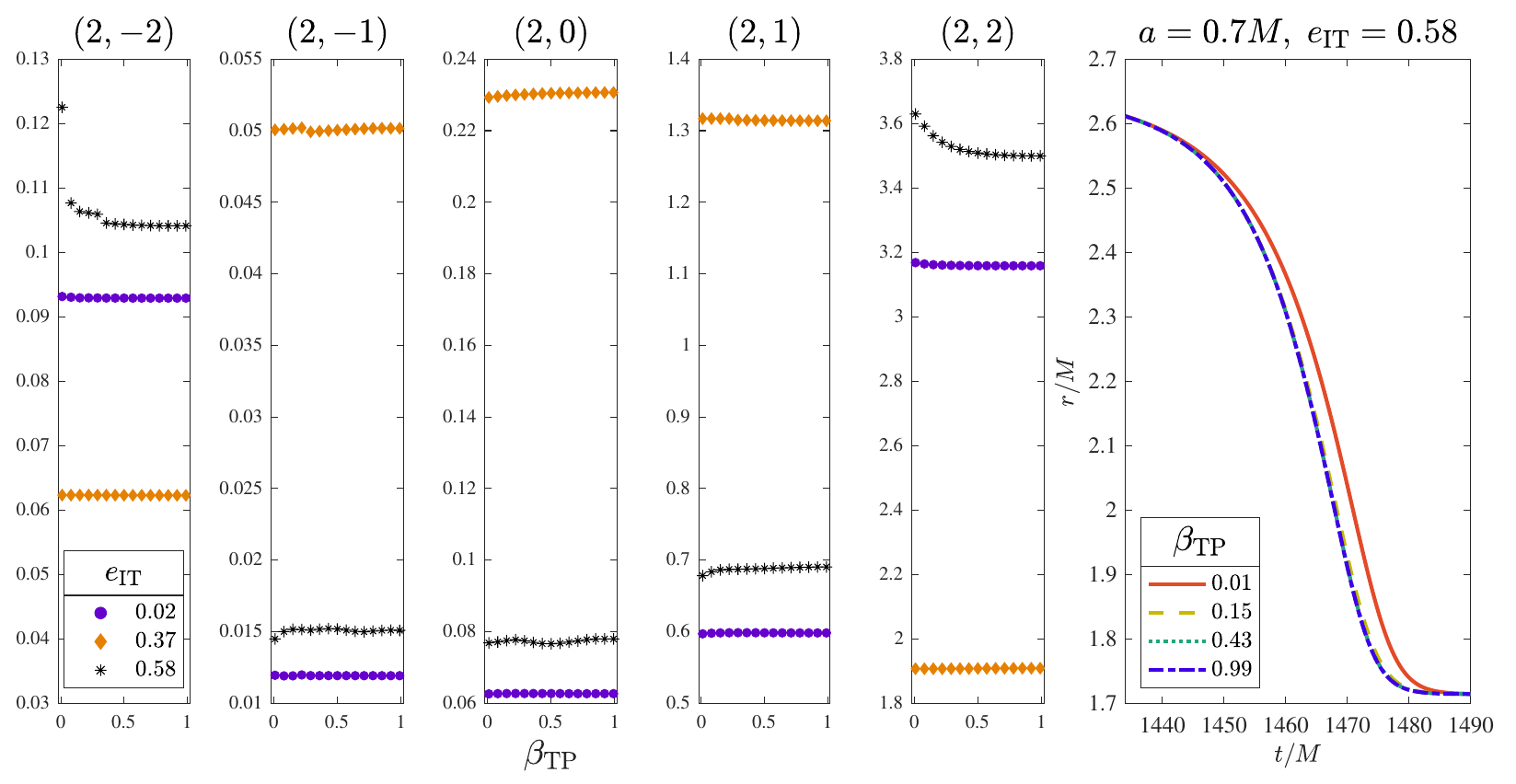}
    \caption{Mode excitation magnitude $\mathcal{A}_{km0}$ for spheroidal QNMs with $k = 2, \;m\in \{-2, -1, 0, 1, 2\}$, $n=0$ as a function of $\beta_{\rm TP}$. All systems have $a = 0.7M$, $\eta = 10^{-4}$ and $e_{\rm IT}\in [0.02, 0.58]$.  The first five panels show the mode amplitudes, and the right-most panel includes four representative worldlines with different values of $\beta_{\rm TP}$. }
    \label{fig:bTP}
\end{figure*}

Figure \ref{fig:bTP} examines the $k = 2$ QNM modes as a function of $\beta_{\rm TP}$.  The left five panels again include the absolute mode amplitudes for $m \in\{-2, ..., 2\}$, and the right-most panel shows representative worldlines that differ only by $\beta_{\rm TP}$.  As in Fig.\ \ref{fig:aIT}, each system has $a = 0.7M$ and $\eta = 10^{-4}$, but now $e_{\rm IT}\in [0.02, 0.58]$.  We see even less variation in the mode amplitudes here than we see when we vary $\alpha_{\rm IT}$, especially if we focus on the range $0.33 \le \beta_{\rm TP} \le 0.99$ advocated for in BH25.  In many cases, the variation cannot be discerned at the resolution of this graphic.  We conclude that our ringdown model is also quite stable against changes in $\beta_{\rm TP}$.

We do not investigate the dependence of power-law tails on our model parameter choices.  Such a study would be computationally expensive.  Given the results we find for QNMs, there does not appear to be a significant need for such an investigation.


\bibliographystyle{unsrt}

\bibliography{EccTransQNM}

\end{document}